\documentclass[pdflatex,sn-mathphys-num]{sn-jnl}

\usepackage{graphicx}%
\usepackage{multirow}%
\usepackage{amsmath,amssymb,amsfonts}%
\usepackage{amsthm}%
\usepackage{mathrsfs}%
\usepackage[title]{appendix}%
\usepackage{xcolor}%
\usepackage{textcomp}%
\usepackage{manyfoot}%
\usepackage{booktabs}%
\usepackage{algorithm}%
\usepackage{algorithmicx}%
\usepackage{algpseudocode}%
\usepackage{listings}%
\usepackage{svg}%
\usepackage{paralist}%
\usepackage{array}%
\usepackage{longtable}%
\usepackage{placeins}%
\usepackage{tabularx}
\usepackage{float}
\usepackage{xspace}
\usepackage{cleveref} 
\usepackage{enumitem} 

\renewcommand{\arraystretch}{1.5}

\theoremstyle{thmstyleone}%
\theoremstyle{thmstyletwo}%

\theoremstyle{thmstylethree}%
\makeatletter
\let\old@sverb\@sverb
\def\@sverb#1{\old@sverb{#1}\zz}
\def\zz#1{#1\ifx\@undefined#1\else\penalty\z@\expandafter\zz\fi}
\makeatother

\newcommand{\token}[1]{%
    \textlangle\texttt{#1}\textrangle\allowbreak
}

\newcommand{\metatoken}[1]{%
\textcolor[HTML]{c66aba}{\texttt{#1}}\allowbreak%
}

\newcommand{\globalent}[1]{%
\textcolor{black}{#1}%
}

\newcommand{\complex}[1]{%
\textcolor{black}{#1}%
}

\newcommand{\molecule}[1]{%
\textcolor{black}{#1}%
}

\newcommand{\sequenceWrp}[1]{%
\textcolor{orange}{
\textbf{#1}}
}

\newcommand{\highlight}[1]{%
\textcolor{red}{#1}%
}

\newcommand{\betachain}{$\beta$-chain\,}
\newcommand{\alphachain}{$\alpha$-chain\,}
\newcommand{\tasktitan}{TCR Bind}
\newcommand{\taskher}{AbAg Bind}
\newcommand{\taskddg}{PPI \ensuremath{\Delta \Delta G}}

\newcolumntype{L}[1]{>{\raggedright\arraybackslash}p{#1}}
\newcolumntype{R}[1]{>{\raggedleft\arraybackslash}p{#1}}

\newcolumntype{M}[1]{>{\centering\arraybackslash}m{#1}}

\newcommand{\MAMMALMODEL}{\textbf{\texttt{ibm/biomed\allowbreak.omics\allowbreak.bl\allowbreak.sm\allowbreak.ma-ted-458m}}\xspace}

\newcommand{\MAMMAL}{MAMMAL }

\begin{document}
\title[Article Title]{MAMMAL - Molecular Aligned Multi-Modal Architecture and Language for Biomedical Discovery }


\author*[1]{\fnm{Yoel} \sur{Shoshan}}\email{yoels@il.ibm.com}
\equalcont{These authors contributed equally to this work.}
\author[1]{\fnm{Moshiko} \sur{Raboh}}
\equalcont{These authors contributed equally to this work.}

\author[1]{\fnm{Michal} \sur{Ozery-Flato}}
\equalcont{These authors contributed equally to this work.}

\author[1]{\fnm{Vadim} \sur{Ratner}}

\author[1]{\fnm{Alex} \sur{Golts}}

\author[2]{\fnm{Jeffrey} \sur{K. Weber}}

\author[1]{\fnm{Ella} \sur{Barkan}}

\author[1]{\fnm{Simona} \sur{Rabinovici-Cohen}}
\author[1]{\fnm{Sagi} \sur{Polaczek}}
\author[1]{\fnm{Ido} \sur{Amos}}
\author[1]{\fnm{Ben} \sur{Shapira}}
\author[1]{\fnm{Liam} \sur{Hazan}}
\author[1]{\fnm{Matan} \sur{Ninio}}
\author[1]{\fnm{Sivan} \sur{Ravid}}
\author[1]{\fnm{Michael M.} \sur{Danziger}}
\author[3]{\fnm{Yosi} \sur{Shamay}}
\author[3]{\fnm{Sharon} \sur{Kurant}}
\author[2]{\fnm{Joseph A.} \sur{Morrone}}
\author[2]{\fnm{Parthasarathy} \sur{Suryanarayanan}}
\author[1]{\fnm{Michal} \sur{Rosen-Zvi}}
\equalcont{These authors contributed equally to this work.}
\author[1]{\fnm{Efrat} \sur{Hexter}}
\equalcont{These authors contributed equally to this work.}

\affil[1]{\orgdiv{IBM Research-Israel}, \orgname{IBM Research}, \orgaddress{ \city{Haifa}, \postcode{3498825},  \country{Israel}}}

\affil[2]{\orgdiv{IBM TJ Watson Research Center}, \orgname{IBM Research}, \orgaddress{\street{1101 Kitchawan Rd.}, \city{NY}, \postcode{10598}, \state{Yorktown Heights}, \country{USA}}}

\affil[3]{\orgdiv{Faculty of Biomedical Engineering}, \orgname{Technion - IIT}, \orgaddress{3200003}, \city{Haifa},  \country{Israel}}

\abstract{
Large language models applied to vast biological datasets have the potential to transform biology by uncovering disease mechanisms and accelerating drug development. However, current models are often siloed, trained separately on small-molecules, proteins, or transcriptomic data, limiting their ability to capture complex, multi-modal interactions.  Effective drug discovery requires computational tools that integrate multiple biological entities while supporting prediction and generation, a challenge existing models struggle to address. For this purpose, we present \textbf{\MAMMAL} -  \textbf{M}olecular \textbf{A}ligned \textbf{M}ulti-\textbf{M}odal \textbf{A}rchitecture and \textbf{L}anguage - a versatile method applied to create a multi-task foundation model that learns from large-scale biological datasets across diverse modalities, including proteins, small-molecules, and omics. MAMMAL’s structured prompt syntax supports classification, regression, and generation tasks while handling token and scalar inputs/outputs.
Evaluated on eleven diverse downstream tasks, it reaches a new state of the art (SOTA) in nine tasks and is comparable to SOTA in two tasks, all within a unified architecture, unlike prior task-specific models. Additionally, we explored Alphafold 3 binding prediction capabilities on antibody-antigen and nanobody–antigen complexes showing significantly better classification performance of MAMMAL in 3 out of 4 targets.
}
\maketitle

\begin{figure}[h!]
\centering


\includegraphics[width = 350pt]{figure_1_full_v_11_svg-raw.pdf}
\caption{ \textbf{(A)} We introduce a multi-align model pretrained on six datasets, each containing tens to hundreds of millions of data points. These data points include protein sequences, small molecules, and gene expression profiles, with a combined sample size of 2 billion. \textbf{(B)} The multi-align model combines flexible encoder-only and encoder-decoder components. It takes sequences as input, which may contain any combination of tokens and scalar elements, processed by an encoder stack consisting of self-attention blocks. In encoder-only mode, a dedicated token prediction head outputs logits for token predictions, with an optional scalar prediction head for scalar outputs. In encoder-decoder mode, residual connections inject features from the encoder’s final hidden layer into each decoder layer, and a decoder-specific prediction head outputs the final logits. \textbf{(C)} Diverse downstream tasks performed by the multi-align model, mapped to their contributions within the steps of a typical drug discovery pipeline. \textbf{(D)} Diverse downstream tasks performed by the multi-align model, categorized by data type used in the fine-tuning process. \textbf{(E)} Performance of the multi-align model across a diverse set of tasks compared to SOTA.      }\label{main_figure}
\end{figure}

\begin{figure}[h!]
\centering

\includegraphics[width = 450pt]{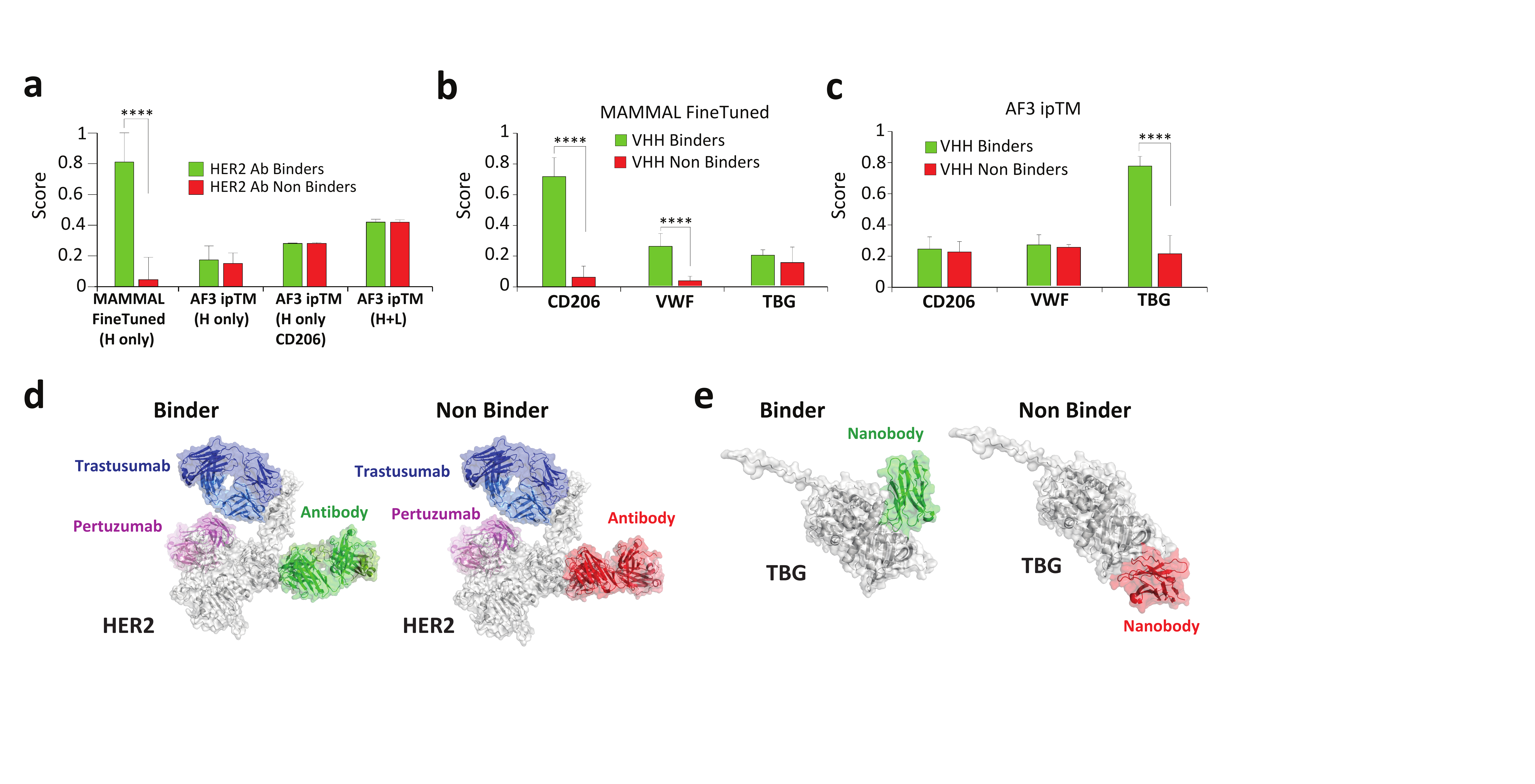}

\caption{ \textbf{Comparison of antibody/nanobody--antigen binding prediction between MAMMAL and AlphaFold~3 (AF3).} \textbf{(a)} HER2 extracellular domain (ECD) binder/non-binder discrimination scores for MAMMAL versus AF3. \textbf{(b-c)} Binding prediction performance for nanobodies (VHHs) against three antigens: CD206, VWF, and TBG (green = binders, red = non-binders; mean ± SD; unpaired two-sided Student's t-test; *$P < 0.05$, ****$P < 0.0001$). \textbf{(b)} MAMMAL evaluation. \textbf{(c)} AF3 ipTM (Interface Predicted TM-score) evaluation. \textbf{(d)} HER2 ECD structure with representative binder/non-binder predictions. Therapeutic antibodies Trastuzumab (blue) and Pertuzumab (purple) shown for comparison. AF3 predicts both binders and non-binders interacting with the same domain, distinct from known therapeutic epitopes.  \textbf{(e)} TBG structure with binding/non-binding VHHs. AF3 predicts distinct poses for binders vs non-binders. 
 }\label{mammal_af3_figure}
\end{figure}

Despite major advances in understanding health and disease, many complex biological processes remain poorly understood. Recent advances in artificial intelligence have given hope that longstanding health challenges may prove treatable with AI-based methods ~\cite{cheng2024artificial} and that the high failure rate of drugs in clinical trials, where approximately $90\%$ of drug candidates do not achieve approval \cite{mullard2016parsing}, can be improved. Over the past century, the process of translating biological discoveries into effective treatments has matured. The drug discovery journey follows a multi-step pipeline that begins with identifying disease-associated proteins, progresses to finding compounds that can effectively target these proteins, and culminates in optimizing drug candidates to meet rigorous standards for efficacy and safety. This process is both costly and labor-intensive, requiring extensive laboratory assays that measure drug-target interactions, assess cellular changes in disease-relevant cell lines, and validate therapeutic efficacy and safety \cite{paul2010improve,dimasi2016innovation,wouters2020estimated}. Drugs in development can be either small molecules, which are stable, easy to manufacture, and suitable for oral delivery \cite{southey2023introduction}, or biologic therapeutics, such as 
engineered antibodies, which offer high specificity but require complex manufacturing and are typically administered by injection \cite{lu2020development}.

Accelerating drug discovery and improving the accuracy of its steps have become a central focus in biomedical research, with the goal of streamlining target identification, drug design, and testing \cite{sadybekov2023computational,huang2024ai,son2024revolutionizing}. Analyzing gene expression profiles, particularly from single-cell RNA sequencing (scRNA-seq), has emerged as a key tool for distinguishing between cell populations associated with different diseases \cite{baslan2017unravelling,ofengeim2017single,rozenblatt2017human,potter2018single}. This analysis enhances our understanding of disease mechanisms, facilitates the identification of new drug targets, and allows for the examination of drug effects across various cell types \cite{van2023applications,Dann2024.04.04.24305313}. In drug design, trained generative models are utilized to synthesize new drug candidates for further exploration \cite{tang2024survey,shanehsazzadeh2023unlocking,swanson2024generative}. As high-throughput screening assays for measuring drug binding affinity are costly and challenging to scale, accurately predicting drug-target interactions can significantly enhance drug design, improving both efficacy and precision. Overall, predictive modeling of binding affinity, toxicity, and efficacy in the early stages of the pipeline can reduce reliance on expensive late-stage testing, ultimately saving time and resources in drug development. 

When creating predictive and generative AI models, one of the key challenges in the field is determining how different modalities and entities should be combined as inputs and outputs to a model \cite{athaya2023multimodal}. This presents a significant hurdle in predictive and generative tasks that involve interactions between entities, for example, predicting whether a specific small-molecule drug and a protein target are likely to bind. While the representation of small molecules and proteins as sequences, such as SMILES and amino acid chains, has demonstrated its value over the years, the approach and benefits of representing transcriptomic data as a sequence have only recently been studied, leading to multiple proposals \cite{szalata2024transformers}. 

In this paper, we propose a cross-domain model that leverages sequence representation for three modalities: small molecules, proteins, and transcriptomic data. Specifically, we introduce MAMMAL (Molecular Aligned Multi-Modal Architecture and Language), a foundational model designed to address the challenges associated with drug discovery tasks. Our key contributions are: (1) a novel multi-align method that combines data from small molecules, proteins, and gene expression using a flexible, structured prompt syntax. This approach enables the model to handle the complex, multi-entity interactions critical for drug discovery tasks such as binding prediction; (2) an architecture specifically designed for drug discovery, featuring seamless integration of numerical values directly into its embedding space via
a projection layer. This improves precision and avoids the limitations of alternative methods, making it ideal for tasks such as predicting binding affinities or the integration of a 3D structure. Additionally, the model’s hybrid design optimizes performance across classification, regression, and generation tasks, all essential in different stages of drug development; and (3) A pre-trained model trained on over 2 billion samples across tasks like spans infilling, denoising, and binding prediction. By achieving SOTA results on 9 out of 11 benchmarks, MAMMAL demonstrates its ability to streamline key steps in the drug discovery pipeline. The model and pre-trained weights are publicly available on \url{https://huggingface.co/ibm/biomed.omics.bl.sm.ma-ted-458m} under the name \MAMMALMODEL, providing a platform for advancing research and innovation in biomedicine.

\section{Results}\label{subsec_results_long}

To evaluate the performance and generalization capabilities of \MAMMALMODEL, we selected a diverse set of existing benchmarks spanning multiple task types and stages of the drug discovery pipeline, prioritizing benchmarks with clearly defined splits when those were available. We assessed model quality through a fine-tuning-based evaluation strategy, where the pretrained model is adapted to each benchmark and compared against specialized state-of-the-art (SOTA) models. The evaluation methodology and fine-tuning protocol as well as detailed descriptions of each benchmark—including background, significance for drug discovery, prior models, and data statistics—are provided in the subsections below. A summary of performance across tasks is presented in Table~\ref{table_main_results} and visualized in Figure~\ref{main_figure}(E), and representative encoder-decoder examples are shown in Table~\ref{table_task_queries}.

AlphaFold~\cite{alphafold}, whose development contributed to the 2024 Nobel Prize in Chemistry, revolutionized protein structure prediction. Its extension AlphaFold-Multimer~\cite{alphafold-multimer} enabled modeling of antibody-antigen complexes, while AlphaFold~3 (AF3)~\cite{alphafold3} further improved accuracy and added nucleic acid/small molecule support. Motivated by AF3's reported advances, we evaluated its performance on therapeutic Antibody and Nanobody complexes (Subsection~\ref{compare_to_af3}). Comparative analysis showed MAMMAL achieved better classification performance than AF3 in three of four targets (Figure~\ref{mammal_af3_figure}).

\begin{table}
\caption{Comparison of SOTA and MAMMAL Performance Across Benchmarks}
\label{table_main_results}
\begin{tabular}{L{2cm} l L{0.8cm} L{1.8cm} L{2cm} L{2cm} r}
\toprule
Benchmark & Domain & Type & Metric & SOTA & MAMMAL & Imp. \\
\midrule

Cell type & GE & cls & $\uparrow$ F1 &  \cite{xu2023ciform} 0.710     & 0.763±0.012 & \textbf{7.5} \% \\    
BBBP & SM & cls & $\uparrow$ AUROC &  \cite{ross2022large} 0.937     & 0.957±0.006 & \textbf{2.2} \% \\       
ClinTox & SM & cls & $\uparrow$ AUROC & \cite{ross2022large} 0.948      & 0.986±0.007 & \textbf{4.0} \% \\   
 
Cancer-Drug Response 1 & GE+SM & reg & $\uparrow$ Pearson & \,\,\,\cite{chaves2024tx} 0.887      & 0.917±0.001 & \textbf{3.4} \% \\

Cancer-Drug Response 2 & GE+SM & reg & $\uparrow$ Pearson & \,\,\,\cite{chaves2024tx} 0.900      & 0.931±0.002 & \textbf{3.4} \% \\

Cancer-Drug Response 3 & GE+SM & reg & $\uparrow$ Pearson & \cite{liu2020deepcdr} 0.923 [0.917-0.929]\  & 0.928±0.000 & 0.5 \% \\
Ab Infilling & Protein & gen & $\uparrow$ CDRH3-AAR & \cite{kong2023end} 0.375      & 0.446±0.002 & \textbf{19.0} \% \\     

\taskher & Protein & cls & $\uparrow$ AUROC & \cite{jing2024accurate} 0.924 [0.923-0.925]  & 0.928±0.002 & 0.4 \% \\

\tasktitan & Protein & cls & $\uparrow$ AUROC & \cite{weber2021titan} 0.862 [0.85-0.868]\  & 0.879±0.003 & \textbf{2.0} \% \\

\taskddg & Protein & reg & $\uparrow$ Pearson & \cite{jin2024attabseq} 0.663      & 0.852±0.041 & \textbf{28.5} \% \\

DTI & Prot.+SM & reg & $\downarrow$ NRMSE & 0.942±0.028 \cite{xu2022peer} & 0.906±0.011 & \textbf{3.8} \% \\

\bottomrule
\end{tabular}
\footnotetext{ For NRMSE, lower is better. For other metrics (AUROC, CDRH3-AAR, Pearson, Spearman, and F1), higher is better.
Each row shows the results from a MAMMAL model fine-tuned from \MAMMALMODEL for the corresponding task.
Abbreviations: In the "Type" column: "cls" = classification; "reg" = regression; "gen" = generation; "Imp." = improvement (percentage) of our model over SOTA. In the "Domain" column: "GE" = genes expression; "SM" = small molecule; "Prot." = protein.
}
\end{table}

\subsection{Evaluation}\label{subsec_eval} 

We compiled a comprehensive set of 11 benchmarks covering multiple data domains and task types, including classification, regression and generation, as well as single-entity, multi-entity, and multi-domain tasks. These benchmarks address key stages of the drug discovery process: Identifying target cell types (Cell Type) and advancing precision medicine (Cancer-Drug Response 1-3); predicting drug efficacy (BBBP) and safety (ClinTox); predicting the binding affinity of small-molecule drugs to target proteins (DTI); predicting interactions of biological drugs (PPI); and designing new drugs, such as antibodies, to target specific proteins (Ab Infilling).  A key criterion in selecting benchmarks was the availability of predefined train, validation, and test splits. For benchmarks with train-validation-test splits, we fine-tuned the model \MAMMALMODEL on the training set, selected the best checkpoint based on validation performance, and reported final results on the test set. Unless otherwise noted, standard errors were estimated by training the models with three different random seeds and calculating the standard deviation of their performance on the test set. Detailed descriptions of each benchmark and the fine-tuning methods used to adapt our pre-trained model for these tasks are provided alongside the evaluation results for each benchmark. In one of the benchmarks (DTI), we report performance using the normalized root mean square error (NRMSE), calculated by dividing the root mean square error by the standard deviation of the labels in the test set. We consider our models to outperform the existing SOTA when the improvement in performance, measured by $|\text{SOTA} - \text{MAMMAL}| \mathbin{/} \text{SOTA}$, exceeds 1\%.

\subsection{Cell Type Annotation}\label{subsec_cta}


Cell type prediction enables researchers to distinguish between different cell populations, such as those associated with various diseases \cite{baslan2017unravelling,ofengeim2017single,rozenblatt2017human,potter2018single}. It is also crucial for understanding how diseases or drugs affect different cell types. In recent years, a variety of methods have been developed for this task, including approaches based on marker genes, correlation-based techniques, and annotation using classification \cite{qi2020clustering}. Recent advances in transformer-based and large-scale foundation models \cite{cui2024scgpt,xu2023ciform,yang2022scbert} have shown improved performance.

The input for this task is single-cell gene expression data. The benchmark we used is based on the Zheng68k dataset \cite{zheng2017massively}, which is composed of human peripheral blood mononuclear cells and is widely used for evaluating cell-type annotation performance, due to the similarity of the cell types involved. The dataset contains 68,579 cells across 11 cell types and originally included 32,738 genes, which after removing non-expressed genes leaves 20,387 genes in the benchmark.  Preprocessing involved normalization, log transformation of expression values, followed by binning. Similar to the approach in \cite{genformer}, our model uses a ranked list of expressed gene names, ordered by their expression levels, as input.  The label to predict is provided in the cell ontology format “CL:NNNNNN” (see Table~\ref{table_task_queries}). 

The model \MAMMALMODEL was fine-tuned and evaluated using 5-fold cross-validation, while ensuring similar proportions of cell types across the folds. As shown in Tables~\ref{table_main_results} and \ref{table_cell_type_results}, MAMMAL outperforms the previous state-of-the-art performance in both accuracy and F1, achieving a 7.5\% improvement in F1.

\subsection{BBBP and ClinTox}
To ensure the development of safe and effective drugs, candidates must satisfy rigorous criteria related to both efficacy and safety. In this study, we selected two relevant benchmarks from MoleculeNet \cite{wu2018moleculenet}, a widely used suite of benchmarks for evaluating machine learning models on small-molecule drug properties: BBBP and ClinTox. The BBBP benchmark focuses on predicting the ability of drugs to penetrate the blood-brain barrier, a critical consideration for drugs targeting the central nervous system. The ClinTox benchmark comprises two related tasks: (1) predicting failure in clinical toxicity trials, and (2) predicting FDA approval status. The overall performance on ClinTox is reported as the average performance across these two tasks.

MoLFormer \cite{ross2022large}, a well-established model for molecular embeddings trained on 1.1 billion SMILES sequences, has achieved state-of-the-art performance on both the BBBP and ClinTox benchmarks. In our study, we adopted the benchmarks from \cite{ross2022large}, which provided predefined splits for training, validation, and testing. As shown in Table
~\ref{table_main_results}, MAMMAL surpassed MoLFormer on both benchmarks, achieving an average area under the receiver operating characteristic curve (AUROC) score of 0.957 on BBBP and 0.986 on ClinTox, representing improvements of 2.2\% and 4\%, respectively, over the state of the art.

\subsection{Cancer-Drug Response}\label{subsec_gdsc}

Identifying drug response at the cellular level is a critical step in the development of new drugs. Two key public databases supporting this effort, particularly in cancer drug development, are the Cancer Cell Line Encyclopedia (CCLE) \cite{barretina2012cancer} and the Genomics of Drug Sensitivity in Cancer (GDSC) \cite{yang2012genomics}. CCLE provides multi-omics profiles for around 1,000 cancer cell lines, while GDSC offers data on the drug responses of these lines to hundreds of drugs, commonly measured using the half-maximal inhibitory concentration (IC$_{50}$). Notable computational models addressed this task \cite{lind2019predicting,liu2020deepcdr,liu2022graphcdr}.

For our study, we used three subsets of the GDSC database: GDSC1 and GDSC2, available through the Therapeutics Data Commons (TDC) \cite{huang2021therapeutics}, and referred to in the paper as Cancer-Drug Response 1 and Cancer-Drug Response 2, respectively; and a subset published in \cite{liu2020deepcdr}, referred to as Cancer-Drug Response 3. Table~\ref{table_gdsc_data_stats} summarizes the number of cell lines, drugs, and cell-drug pairs in these datasets. We used the random splits provided by TDC for Cancer-Drug Response 1 and 2, while for Cancer-Drug Response 3, we followed the split methodology outlined in \cite{liu2020deepcdr}, reserving 5\% of the data for the test set, stratified by TCGA \cite{weinstein2013cancer} pathways associated with the cancer cell lines.

During fine-tuning, we used only gene-expression profiles and SMILES representations of drugs, as shown in the example prompt in Table~\ref{table_task_queries}. Similar to the input format for cell type annotation, gene-expression profiles were provided as ranked lists of gene names based on their expression levels. For predicting continuous IC$_{50}$ values, MAMMAL was utilized in regression mode, taking advantage of its built-in support for floating-point scalar predictions. As demonstrated in Table~\ref{table_main_results}, our model outperforms the current SOTA models for Cancer-Drug Response 1 and 2, achieving a 3.4\% increase in Pearson correlation values. Additionally, it yields results comparable to the SOTA for the Cancer-Drug Response 3 benchmark, with a slight improvement of 0.5\%.

\subsection{Targeted Antibody Design}\label{subsec_antibody_design}
Antibodies are a family of proteins produced by the immune system to neutralize foreign antigens and are of particular interest due to their high specificity and strong binding to target molecules \cite{hummer2022advances, chiu2019antibody}. These characteristics have made them a crucial class of therapeutics, driving significant research efforts into the design of new antibody-based drug candidates \cite{basu2019recombinant, carter2018next, beck2017strategies, lu2020development}.  Antigen-binding fragments (Fabs) are the antibody fragments that bind to antigens. It is composed of one constant  domain and one variable domain of each of the \textit{heavy} and \textit{light} chains. Each variable region is further divided into four framework (FR) regions and three complementarity-determining regions (CDRs). While FR regions are typically conserved, CDRs exhibit significant variation in their amino acid composition and are generally the primary determinants of binding affinity to the target antigen. When designing novel antibodies for a specific antigen, the typical approach is to explore alternative CDRs that could produce a new, functional antibody with high binding affinity to the target \cite{hummer2022advances, chiu2019antibody, saka2021antibody, kong2023end}.

Recently, several deep learning methods have been developed for targeted antibody design, framing CDR prediction as an \textit{infilling} task \cite{saka2021antibody, jin2021iterative, jin2022antibody, luo2022antigen, kong2022conditional, kong2023end, zhou2024antigen}. These models predict missing CDR regions, represented by \textit{MASK} tokens, using the amino acid sequences of both the antigen and the antibody's FR regions. While prior approaches often relied on structural data, this information is scarce and challenging to obtain \cite{dunbar2014sabdab}. In contrast, we fine-tune MAMMAL for the targeted antibody design task using only the sequence data of the antigen and the sequence of the antibody's FR regions. 

The targeted antibody design task benchmark is based on the SAbDab dataset \cite{dunbar2014sabdab}. Following the data processing outlined in \cite{kong2023end}, we filtered out samples with missing CDRs to enable direct comparison, even though MAMMAL supports samples that contain missing CDRs. Consistent with \cite{kong2023end}, we randomly partitioned the dataset into training, validation, and test folds while ensuring that samples with similar heavy CDR3 sub-sequences remained in the same fold. As demonstrated in Tables~\ref{table_main_results} and \ref{table_antibody_design_results}, MAMMAL shows superior amino acid recovery across all masked CDRs. Notably, in CDRH3, the most variable region, it exhibits a remarkable improvement of 19\%.

\subsection{T-Cell Receptor-Epitope Binding}
T-cell receptor (TCR) binding to immunogenic peptides (epitopes) presented by major histocompatibility complex molecules is a critical mechanism in the adaptive immune system, essential for antigen recognition and triggering immune responses. 
The TCR repertoire exhibits considerable diversity, consisting of an \alphachain and a \betachain that function together to enable T cells to recognize a wide array of epitopes. The \betachain is especially significant, as it is crucial for the early stages of T-cell development and possesses greater variability, which enhances the TCR's capacity to identify diverse pathogens effectively. 
However, understanding the specific interactions between TCRs and epitopes remains a significant challenge due to the vast variability in TCR sequences. 
Accurate prediction of TCR-peptide binding from sequence data could revolutionize immunology by offering deeper insights into a patient's immune status and disease history. 
This capability holds potential applications in personalized immunotherapy, early diagnosis, and the treatment of diseases such as cancer and autoimmune disorders. 
In silico tools designed to model TCR-peptide interactions could also facilitate the study of therapeutic T-cell efficacy and assess cross-reactivity risks, presenting a transformative opportunity for precision medicine.

We evaluated the model on the task of predicting TCR-epitope binding from sequence data using the Weber benchmark (\cite{weber2021titan}, \cite{noauthor_tcr-epitope_nodate}), which consists of 47,182 TCR {\betachain} epitope pairs. This dataset covers 192 distinct epitopes and includes 23,139 unique TCR \betachain sequences, with 50\% of the pairs serving as negative samples created by randomly pairing TCR sequences with epitopes they are not known to bind with. The dataset also includes the CDR3 subsequence for each TCR \betachain, the most hypervariable region of the chain. We used 10-fold cross-validation. The folds were pre-defined in   \cite{weber2021titan}. Fine-tuning involved three concurrent tasks: TCR \betachain mask infilling and two classification tasks: (i) TCR \betachain epitope binding prediction and (ii) TCR \betachain-CDR3 epitope binding prediction. Here, we report the performance only for the TCR \betachain epitope binding prediction task. As depicted in Table \ref{table_main_results}, our model achieved an average AUROC of 0.879, representing a statistically significant improvement of 2\% over the SOTA, as our result falls outside the SOTA's confidence interval.

\subsection{Protein-Protein Interaction - $\Delta$$\Delta$G Prediction}\label{subsec_ddg}

An important factor in drug design is binding affinity, commonly measured by the equilibrium dissociation constant, $K_{D}$, which is related to the Gibbs free energy $\Delta{G}$ through the equation
\label{eq:ddg}
\begin{equation}
    \Delta G = kT \ln(K_{D}),
\end{equation}
where $k$ is the Boltzmann constant and $T$ is the temperature \cite{jankauskaite2019skempi}.  

The effect of mutating several residues in a protein complex on binding affinity can be quantified by the difference in $\Delta G$ between the mutant and the reference (wild-type) complex. This difference is expressed as $$\Delta \Delta G = \Delta G_{\text{mutant}} - \Delta G_{\text{wild-type}}.$$ Predicting $\Delta \Delta G$ is a central focus of numerous research efforts \cite{liu2023persistent, wang2020topology, guo2022machine}.

The SKEMPI dataset \cite{jankauskaite2019skempi} provides information on changes in thermodynamic parameters, including $\Delta G$, and kinetic rate constants due to mutations in protein-protein (PP) complexes whose structures are available in the Protein Data Bank \cite{10.1093/nar/28.1.235}.
This dataset is extensively utilized in the literature for predicting the effects of mutations on binding affinity, particularly in the context of $\Delta \Delta G$.
A subset of SKEMPI comprising 1,131 samples of single mutations, S1131, is widely used. We adopt S1131 as our benchmark for predicting protein-protein $\Delta \Delta G$ and follow the common practice of reporting 10-fold cross-validation performance on this subset. The input prompt for our model includes the reference ``wild-type" version of the complex and the corresponding mutated version, comprising only sequence data without any structural information. We leverage MAMMAL's support for floating-point scalars to predict continuous $\Delta \Delta G$ in a regression task setting. Performance results are presented in Table~\ref{table_main_results}. As shown in Table~\ref{table_main_results}, our model achieved an average Pearson correlation of 0.852, significantly exceeding the previous sequence-only SOTA of 0.663. Compared to models that incorporate structural data, our model's performance remains competitive, falling just 1.6\% short of the SOTA performance of 0.866 \cite{liu2023persistent}.

\subsection{Drug-Target Interaction}\label{subsec_dti} 
 Predicting drug-target binding affinity plays a crucial role in the early stages of drug discovery. Traditionally, binding affinities are measured through high-throughput screening experiments, which, while accurate, are resource-intensive and limited in their scalability to evaluate large sets of drug candidates. In this task, we focus on predicting binding affinities using $\textnormal{p}K_{D}$, the negative logarithm of the dissociation constant, which reflects the strength of the interaction between a small molecule (drug) and a protein (target). We utilize the PEER (Protein sEquence undERstanding) benchmark \cite{xu2022peer} for DTI prediction. This benchmark leverages data from the BindingDB dataset \cite{bindingdb}, with a specific test split that holds out four protein classes - estrogen receptor, G-protein-coupled receptors, ion channels, and receptor tyrosine kinases - for assessing generalization performance on unseen classes.

For model fine-tuning, we conducted hyperparameter optimization, selecting an initial learning rate of 0.0004, with no dropout and no weight decay. We standardized the $\textnormal{p}K_{D}$ values based on the mean and standard deviation of the training set. For evaluation, we transformed the predicted values back to their original scale. As shown in Table \ref{table_main_results}, our model achieved an average NRMSE of 0.906, demonstrating a solid improvement of 3.8\% over the SOTA reported by \cite{xu2022peer}.

\subsection{Antibody-Antigen Binding Prediction}\label{binding_prediction}
Accurate prediction of antigen-antibody binding can enhance the design and optimization of therapeutic antibodies, leading to improved efficacy and specificity. We employ the human epidermal growth factor receptor 2 (HER2) dataset \cite{mason_optimization_2021} as a benchmark for predicting antibody-antigen binding. HER2 is a key target for certain types of breast and stomach cancers. The dataset includes variations of the clinically approved therapeutic antibody trastuzumab and their corresponding affinities for the HER2 antigen. The dataset comprises 8,935 binding and 25,114 non-binding trastuzumab CDR H3 mutants, each with up to 10 mutations, following de-duplication and the removal of samples labeled as both binding and non-binding. 

The HER2 dataset was divided into train (70\%), validation (15\%) and test (15\%) sets. 

Finetuning involved feeding the target antigen sequence as well as the entire heavy-chain variable region as input and predicting binding to the target sequence. As depicted in Table~\ref{table_main_results}, our model achieved an average ROCAUC of 0.928, slightly surpassing the SOTA, which incorporated structural data, unlike our model.

\subsection{Comparison of AlphaFold 3 and MAMMAL in Predicting Antibody–Antigen and Nanobody–Antigen Binding}\label{compare_to_af3}

Accurate prediction of antibody–antigen and nanobody–antigen interactions is essential for evaluating therapeutic efficacy and guiding protein engineering. Although AlphaFold 3 (AF3) is not explicitly designed as a binary protein–protein interaction (PPI) classifier, it can infer binding likelihood when the predicted PPI structure exhibits a well-packed interface with high confidence, as measured by predicted template modeling (pTM) or interface predicted template modeling (ipTM) scores. Recent studies suggest that these scores correlate with true binding events \cite{Bennett2024.03.14.585103, af_ab_modeling}. We conducted an exploratory comparison of AF3’s ability to distinguish binders from non-binders against a fine-tuned MAMMAL model.

We first evaluated the extracellular domain (ECD) of HER2, a well-characterized therapeutic antigen with known binding epitopes. Figure~\ref{mammal_af3_figure}a compares the score distributions for MAMMAL and AF3 in classifying binders versus non-binders. Due to the computational demands and limited availability of AF3, we downsampled the test set to 60 examples (30 positive and 30 negative). We used the HER2-specific MAMMAL model described in Subsection~\ref{binding_prediction}, which achieved strong discriminative performance (AUROC = $0.93$; $P < 0.0001$, unpaired two-sided $t$-test). Conversely, AF3 exhibited no meaningful separation (AUROC $\approx 0.5$) across all tested configurations. Furthermore, AF3's predicted ECD binding sites differed from the known epitopes of the FDA-approved antibodies trastuzumab and pertuzumab, as visualized in Figure~\ref{mammal_af3_figure}d. Extended analyses are provided in Appendix~\ref{appendix_alphafold3}.

We next evaluated three structurally diverse targets: (i) CD206 (a 1500-amino-acid transmembrane protein), (ii) von Willebrand Factor (VWF; a $>$2000-amino-acid serum protein), and (iii) thyroxine-binding globulin (TBG; a 384-amino-acid serum protein). Binders were nanobodies sourced from SAbDab-nano \cite{sabdab-nano}, patents, and proprietary datasets, while non-binders comprised random nanobodies against unrelated targets and non-binders from a phage display library. From the full dataset (154 sequences: 46 binders and 108 non-binders), we selected 50\% (binders and non-binders) for MAMMAL finetuning and 50\% for evaluation. The fine-tuned MAMMAL model exhibited strong performance on CD206 and VWF (AUROC $=0.995$) and $0.83$ respectively) but performed poorly on TBG (AUROC = $0.7$), as shown in Figure~\ref{mammal_af3_figure}b. In contrast, AF3 showed the opposite trend, achieving strong performance on the smaller TBG target (AUROC = $0.98$) but weak discrimination for the larger CD206 and VWF proteins (AUROC $<0.57$), as shown in Figure~\ref{mammal_af3_figure}c. Structural analyses revealed distinct AF3-predicted binding sites for binders versus non-binders (Figure~\ref{mammal_af3_figure}e). Extended analyses for all targets are provided in Appendix~\ref{appendix_alphafold3}.

\section{Discussion}\label{sec12}
The rapid advancement of biomedical AI foundation models has focused on different aspects of cellular content. These include RNA expression-based models (e.g., \cite{cui2024scgpt, yang2022scbert, genformer}), protein-based models (e.g., \cite{uniprot, madani2020progen}), and models of small molecules interacting with proteins (e.g., \cite{ross2022large}). As a result, we are moving closer to uncovering previously unknown biological processes. We may even be on the brink of solving the fundamental challenge of building a virtual cell. This concept has recently been proposed as a futuristic AI-based tool for novel biological discoveries and the design of new treatments \cite{bunne2024build, song2024toward}. A crucial first step toward this goal is developing methods to integrate these diverse models. The MAMMAL approach proposed in this paper enables such integration. It reformulates tasks as sequence-to-sequence problems and introduces several key architectural enhancements: support for both encoder-only and encoder-decoder modes, a multi-domain extensible syntax for inputs and outputs, and the direct handling of numerical values through continuous token embeddings. A model was  pretrained using this approach  while aligning inputs across diverse datasets --- including small molecule, protein, antibody, and gene expression data --- using a variety of pretraining tasks. This multi-align approach enables the integration of cross-domain pharmaceutical knowledge into a single model and facilitates effective transfer learning to fine-tuned models for downstream applications. Demonstrated state-of-the-art performance of the multi-align fine-tuned models across diverse tasks, spanning multiple data domains and stages of the drug discovery pipeline, firmly establishes the MAMMAL approach as a powerful tool with transformative potential.

We also explored AF3 capabilities in predicting binding. We observed underwhelming performance of AF3 in three out of four cases, which might be a result of a built-in bias in the AF3 training process, such as exposing the model only to true positive binders. Another hypothesis is that sequence-based models are capable of learning from a more varied collection of protein chains, including proteins with Intrinsically Disordered Regions (IDR). Additionally, AF3 was not directly trained on separating binders and non-binders, which can also explain its underwhelming performance on this task. Going forward, we plan to expand MAMMAL to natively support 3D input and output, to enjoy the benefits of both sequence and structure-based methods.

In recent months, the world has rapidly transitioned to embrace agentic workflows, an advanced methodology that leverages large language models (LLMs) and specialized agents to automate and enhance discovery processes traditionally performed by humans \cite{gridach2025agentic,ramos2025review}. This shift marks a significant evolution in how research and development are conducted across industries. The fine-tuned models derived from MAMMAL seamlessly integrate into this paradigm, offering highly specialized capabilities that enhance efficiency and accuracy. By capitalizing on the power of multi-domain learning and tailored fine-tuning, these models are poised to revolutionize workflows, breaking new ground and unlocking unprecedented potential in the field of biomedical discovery.

\backmatter

\section{Methods}\label{subsec_method}
The MAMMAL method is built around three core components: The model architecture, the molecular prompt syntax, and extensive pretraining. In Subsection \ref{subsec_arch}, we detail the architecture and its enhancements to the standard transformer. 

Subsections \ref{entity_representation} and \ref{subsec_quer_sntx} focus on the molecular representations and prompt syntax, key features that enable the support of a diverse range of pretraining and downstream tasks for drug discovery. Finally, Subsection \ref{subsec_pretraining_short} outlines the pretraining process, which facilitates leveraging large, cross-domain datasets and handling multiple entities simultaneously. 

\subsection{MAMMAL Architecture}\label{subsec_arch}

The MAMMAL architecture builds upon the transformer architecture introduced by Vaswani et al. \cite{vaswani2017attention} and draws inspiration from the T5 framework \cite{t5} while introducing several key modifications on top of it. This design conceptualizes tasks as sequence-to-sequence problems within a unified model, introducing three key features:
\\\\
\noindent\textbf{Optimized for Representation and Generation Tasks}

\noindent Biomedical models for drug discovery must perform diverse tasks, ranging from token-level or prompt-level representation to tasks requiring strong generative capabilities. For instance, \cite{XTrimoPGLM} introduced a non-causal, decoder-only protein language model optimized for both understanding and generating protein-related data, outperforming state-of-the-art baselines.
MAMMAL is designed to jointly optimize for such diverse tasks, enabling the model to learn from multiple stages of the drug discovery pipeline. This collective learning is facilitated by a shared encoder, which bridges gaps between distinct AI applications in drug discovery.
Specifically, MAMMAL supports both encoder-only and encoder-decoder autoregressive modes. Encoder-only mode excels in representation-heavy tasks, such as classification and regression, while encoder-decoder mode is better suited for generative tasks. By sharing encoder stack weights across these modes, MAMMAL facilitates efficient multi-task training, with parameter updates conducted through gradient accumulation across all tasks.
\\\\

\vspace*{5px}
\noindent\textbf{Flexible Multi-Domain Structured Prompts}

\noindent Drug discovery tasks often require prompts that describe molecular complexes composed of multiple entities. These prompts may also include attributes related to the entire complex or specific internal entities to improve prediction accuracy.
To address this, MAMMAL employs a modular tokenizer that enables multi-domain structured prompts by assigning distinct sub-tokenizers to each entity type. Compared to free-text prompts, structured prompts provide a consistent and explicit representation of input data, capturing relationships and attributes between entities, and facilitating the training. 
While MAMMAL primarily focuses on structured input, incorporating free text as an additional data source is a promising future direction. Recent advancements in biomedical large language models \cite{pei2023biot5,pei2024biot5+,chaves2024tx}, demonstrate the potential of free-text prompts to enhance versatility and expand training datasets. Future iterations of MAMMAL could integrate free text directly, mix it with structured prompts, or leverage embeddings derived from extensively pre-trained text models \cite{llama,granite}. These embeddings could harness the latent knowledge encoded within external models to enrich the system’s performance.
\\\\

\noindent\textbf{Numerical Values Integration}

\noindent A notable innovation of MAMMAL is its native support for numerical values (e.g., 2, 3.14, 10,000.1) as both inputs and outputs. Numerical data is crucial for modeling drug discovery tasks and allows for direct representation of important attributes. This capability expands the range of tasks MAMMAL can handle, such as binding affinity prediction and protein folding.
Many biomedical large language models either lack support for numerical data \cite{pei2023biot5}, employ limited solutions such as binning \cite{chaves2024tx}, or rely on digit-based representations \cite{pei2024biot5+}, which can inflate input length. \cite{regression_transformers} offer compelling digit-based approach of splitting the number into semantic parts (mainly digits), supporting arbitrary scalar values while adding only a limited number of tokens to the vocabulary. While this simplifies integration with standard language models, it inflates the number of input/output tokens significantly. This is especially evident in tasks like gene expression prediction which require supporting thousands of scalars in a single prompt and expected model output. \cite{regress_dont_guess} suggests a new loss term that considers numerical proximity, with the main advantage of not requiring any additional prediction head for scalars, allowing to introduce numerical values support within existing language model architectures. However, it is currently limited to a fixed vocabulary, not allowing support for arbitrary (possibly unseen in train time) scalar values.  Other methods, such as discretization, demonstrated in \cite{esm3} through pre-trained VQ-VAE \cite{vqvae} that encodes and decodes 3D spatial positions, represent another approach limited to a specific use case.
MAMMAL, however, integrates continuous numerical values directly into its embedding space via a projection layer. The resulting embeddings align with the input token embeddings, ensuring simple, efficient, and effective utilization of numerical data.
\\\\
\noindent Further details about the MAMMAL architecture can be found in Appendix A.

\subsection{Entity Representation}\label{entity_representation}

\noindent Each entity domain supported by MAMMAL makes use of a representation approach that has been selected as follows:
\begin{itemize}[leftmargin=*]
\item \textbf{Small Molecules.} Represented using SMILES sequences (Simplified Molecular Input Line Entry System). For example, paracetamol is represented as {CC(=O)NC1=CC=C(O)C=C1}.
\item \textbf{Gene Expression.} Represented as an ordered list of gene names, sorted in descending order based on log-normalized and binned expression values. In case of ties, genes are sorted alphabetically. This representation applies to both single-cell and bulk RNA expression data.
\item \textbf{Proteins.} Represented as concatenated amino acid chains, preserving original source-data order. This representation contains only sequence data, without structural information.
\item \textbf{Antibodies.} Represented similarly to proteins, using concatenated amino acid sequences. Each chain is prefixed with a token indicating its type (heavy or light).
\end{itemize}

\subsection{Prompt Syntax}\label{subsec_quer_sntx}

The prompt syntax is built around a modular tokenizer architecture that governs how molecular data is interpreted and encoded. At its core, the tokenizer defines a common syntax using special tags that represent molecular entities, sequences, attributes, and interactions across diverse molecular systems. It supports multi-domain inputs by delegating different segments of the input sequence to specialized sub-tokenizers, each responsible for handling a specific data type or modality.

All sub-tokenizers operate under the umbrella of the main tokenizer and share access to a unified set of special tokens, such as $\langle \text{EOS} \rangle$, ensuring consistency across domains. Among these sub-tokenizers, a designated numeric sub-tokenizer is used for handling continuous values. Instead of tokenizing numeric values into discrete tokens, this sub-tokenizer projects them directly into the model’s embedding space via a learned projection layer.

The prompt syntax applies uniformly to both inputs and outputs of the model. It is also designed to be extensible: New tags can be introduced into individual sub-tokenizers or added to the shared token set without disrupting existing functionality. Thanks to this modular structure, updates to domain-specific sub-tokenizers or the token vocabulary maintain backward compatibility, enabling interoperability between newly trained and previously deployed models. More detailed explanations and examples of the prompt syntax can be found in Appendix~\ref{query_syntax_detailed}.

\subsection{Pretraining}\label{subsec_pretraining_short}
MAMMAL is designed as a comprehensive foundation model, capable of spanning multiple domains and accommodating a variety of entities. It is intended to support diverse task types, ranging from representation-focused tasks to generation-oriented ones. To achieve this, MAMMAL is trained on multiple tasks concurrently. Pretraining was conducted on two billion samples sourced from six datasets, which are all publicly available, covering three distinct domains across seven tasks. Table \ref{table_pretraining_tasks} summarizes these tasks, detailing the relevant domains, entity types, and specific datasets. Additional details about the pretraining are provided in Appendix~\ref{appendix_pretraining}.

\begin{table}
\caption{Pretraining Tasks}
\label{table_pretraining_tasks}
\begin{tabular}{L{2.1cm} L{1.1cm} L{1.1cm} L{2.1cm} l L{1.5cm}}
\toprule
Name & Domain & Entity Type & Task Type & Dataset & Number of Samples \\
\midrule
Protein LM	& Biologic & General Protein & Spans Masking LM & Uniref90 \cite{uniprot} & 180M \\
Antibody LM	& Biologic & Antibody & Spans Masking LM & OAS \cite{oas} & 650M \\
Small Molecule LM & Small Molecules	& Small Molecule & Spans Masking LM	& \begin{tabular}[x]{@{}c@{}}ZINC \cite{zinc22} +\\PubChem \cite{pubchem} \end{tabular}  & 200M \\
Cell Genes LM & Single Cell Transcript- omics & Cell Genes & Spans Masking LM & CELLxGENE \cite{CELLxGENE} & 30M \\
Protein-Protein Interaction	& Biologic & General Protein &	Classification & STRING \cite{string} & 780M \\
Protein-Protein Interaction Gen. & Biologic	& General protein & Generation & STRING \cite{string} & 390M \\
Antibody Denoise & Biologic & Antibody & Denoise Sequence & OAS \cite{oas} & 650M \\
\end{tabular}
\footnotetext{ Details on the pretraining tasks that were used while training \MAMMALMODEL .\\ "Number of Samples" lists the post-filtering number of samples actually used. A single model was pretrained with all of the listed tasks, accumulating knowledge spanning multiple domains.
}
\end{table}

\section{Data Availability}
All datasets used in this study are publicly available. \\
\textbf{Cell Type.} The Zheng68k dataset was obtained from \url{https://www.10xgenomics.com/datasets/fresh-68-k-pbm-cs-donor-a-1-standard-1-1-0} (file: \url{https://cf.10xgenomics.com/samples/cell-exp/1.1.0/fresh_68k_pbmc_donor_a/fresh_68k_pbmc_donor_a_filtered_gene_bc_matrices.tar.gz})

\textbf{BBBP and ClinTox} benchmarks were obtained from \url{https://github.com/IBM/molformer/tree/main/data} that points to \url{https://ibm.ent.box.com/v/MoLFormer-data} (file: \texttt{finetune\_datasets.zip}). 

\textbf{Cancer-Drug Response 1 and 2.} The GDSC1 and GDSC2 benchmarks were accessed with random splits from the TDC library (\url{https://pypi.org/project/PyTDC/}). \textbf{Cancer-Drug Response 3.} The benchmark was obtained from the DeepCDR \cite{liu2020deepcdr} git repository (\url{https://github.com/kimmo1019/DeepCDR/tree/master/data}).

\textbf{DTI.} This benchmark was published by \cite{xu2022peer} and is available in \url{https://torchdrug.ai/docs/api/datasets.html#bindingdb}

\textbf{Ab Infilling.} This data was taken from \cite{kong2023end}, which provides a preprocessed subset of the publicly available SAbDab database \cite{dunbar2014sabdab}. The preprocessing pipeline includes a similarity-based clustering for the data splits and sample-level filtering that excludes samples considered invalid in \cite{kong2023end}. For additional information, we refer to \cite{kong2023end} and the publicly available codebase: \url{https://github.com/THUNLP-MT/dyMEAN}.

\textbf{\taskddg.} The SKEMPI S1131 dataset of non-redundant single mutations was derived from SKEMPI \cite{jankauskaite2019skempi} in \cite{xiong2017bindprofx} and can be downloaded from \url{https://zhanggroup.org/BindProfX/download/}.

\textbf{TCR Bind} The Weber TCR binding dataset was downloaded from \cite{noauthor_tcr-epitope_nodate}, and the HER2 antibody-antigen binding dataset was taken from the original paper \cite{mason_optimization_2021} github repository \cite{HER2_repo_2024}.

\textbf{Pretraining}. The \MAMMALMODEL was pre-trained over OAS \cite{oas}, UniProt \cite{uniref}, Zinc \cite{zinc22}, PubChem \cite{pubchem}, STRING \cite{string} and CELLxGENE \cite{CELLxGENE}. Appendix~\ref{appendix_pretraining} describes the pre-processing steps applied.

\section{Code Availability}

The model architecture, fine-tuning framework, and end-to-end examples are publicly available at:
\url{https://github.com/BiomedSciAI/biomed-multi-alignment}.
This GitHub repository provides comprehensive resources for working with the model, including setup instructions, fine-tuning guidelines, and inference workflows for multiple downstream tasks.

Pretrained model weights and the associated tokenizer are hosted on the Hugging Face Model Hub and can be accessed at:
\url{https://huggingface.co/ibm/biomed.omics.bl.sm.ma-ted-458m}.

Additionally, selected fine-tuned model checkpoints and tokenizers are available via:
\url{https://huggingface.co/models?other=base_model:finetune:ibm-research/biomed.omics.bl.sm.ma-ted-458m},
and can be explored interactively through the Hugging Face Space at:
\url{https://huggingface.co/spaces/ibm/biomed-multi-alignment}.

\section{Acknowledgments}

Partially Supported by IBM-Technion Research Collaboration

\bibliography{sn-bibliography}


\begin{thebibliography}{94}
\ifx \bisbn   \undefined \def \bisbn  #1{ISBN #1}\fi
\ifx \binits  \undefined \def \binits#1{#1}\fi
\ifx \bauthor  \undefined \def \bauthor#1{#1}\fi
\ifx \batitle  \undefined \def \batitle#1{#1}\fi
\ifx \bjtitle  \undefined \def \bjtitle#1{#1}\fi
\ifx \bvolume  \undefined \def \bvolume#1{\textbf{#1}}\fi
\ifx \byear  \undefined \def \byear#1{#1}\fi
\ifx \bissue  \undefined \def \bissue#1{#1}\fi
\ifx \bfpage  \undefined \def \bfpage#1{#1}\fi
\ifx \blpage  \undefined \def \blpage #1{#1}\fi
\ifx \burl  \undefined \def \burl#1{\textsf{#1}}\fi
\ifx \doiurl  \undefined \def \doiurl#1{\url{https://doi.org/#1}}\fi
\ifx \betal  \undefined \def \betal{\textit{et al.}}\fi
\ifx \binstitute  \undefined \def \binstitute#1{#1}\fi
\ifx \binstitutionaled  \undefined \def \binstitutionaled#1{#1}\fi
\ifx \bctitle  \undefined \def \bctitle#1{#1}\fi
\ifx \beditor  \undefined \def \beditor#1{#1}\fi
\ifx \bpublisher  \undefined \def \bpublisher#1{#1}\fi
\ifx \bbtitle  \undefined \def \bbtitle#1{#1}\fi
\ifx \bedition  \undefined \def \bedition#1{#1}\fi
\ifx \bseriesno  \undefined \def \bseriesno#1{#1}\fi
\ifx \blocation  \undefined \def \blocation#1{#1}\fi
\ifx \bsertitle  \undefined \def \bsertitle#1{#1}\fi
\ifx \bsnm \undefined \def \bsnm#1{#1}\fi
\ifx \bsuffix \undefined \def \bsuffix#1{#1}\fi
\ifx \bparticle \undefined \def \bparticle#1{#1}\fi
\ifx \barticle \undefined \def \barticle#1{#1}\fi
\bibcommenthead
\ifx \bconfdate \undefined \def \bconfdate #1{#1}\fi
\ifx \botherref \undefined \def \botherref #1{#1}\fi
\ifx \url \undefined \def \url#1{\textsf{#1}}\fi
\ifx \bchapter \undefined \def \bchapter#1{#1}\fi
\ifx \bbook \undefined \def \bbook#1{#1}\fi
\ifx \bcomment \undefined \def \bcomment#1{#1}\fi
\ifx \oauthor \undefined \def \oauthor#1{#1}\fi
\ifx \citeauthoryear \undefined \def \citeauthoryear#1{#1}\fi
\ifx \endbibitem  \undefined \def \endbibitem {}\fi
\ifx \bconflocation  \undefined \def \bconflocation#1{#1}\fi
\ifx \arxivurl  \undefined \def \arxivurl#1{\textsf{#1}}\fi
\csname PreBibitemsHook\endcsname

\bibitem[\protect\citeauthoryear{Cheng et~al.}{2024}]{cheng2024artificial}
\begin{botherref}
\oauthor{\bsnm{Cheng}, \binits{F.}},
\oauthor{\bsnm{Wang}, \binits{F.}},
\oauthor{\bsnm{Tang}, \binits{J.}},
\oauthor{\bsnm{Zhou}, \binits{Y.}},
\oauthor{\bsnm{Fu}, \binits{Z.}},
\oauthor{\bsnm{Zhang}, \binits{P.}},
\oauthor{\bsnm{Haines}, \binits{J.L.}},
\oauthor{\bsnm{Leverenz}, \binits{J.B.}},
\oauthor{\bsnm{Gan}, \binits{L.}},
\oauthor{\bsnm{Hu}, \binits{J.}}, et al.:
Artificial intelligence and open science in discovery of disease-modifying
  medicines for alzheimer’s disease.
Cell Reports Medicine
\textbf{5}(2)
(2024)
\end{botherref}
\endbibitem

\bibitem[\protect\citeauthoryear{Mullard}{2016}]{mullard2016parsing}
\begin{barticle}
\bauthor{\bsnm{Mullard}, \binits{A.}}:
\batitle{Parsing clinical success rates}.
\bjtitle{Nature Reviews Drug Discovery}
\bvolume{15}(\bissue{7}),
\bfpage{447}--\blpage{448}
(\byear{2016})
\end{barticle}
\endbibitem

\bibitem[\protect\citeauthoryear{Paul et~al.}{2010}]{paul2010improve}
\begin{barticle}
\bauthor{\bsnm{Paul}, \binits{S.M.}},
\bauthor{\bsnm{Mytelka}, \binits{D.S.}},
\bauthor{\bsnm{Dunwiddie}, \binits{C.T.}},
\bauthor{\bsnm{Persinger}, \binits{C.C.}},
\bauthor{\bsnm{Munos}, \binits{B.H.}},
\bauthor{\bsnm{Lindborg}, \binits{S.R.}},
\bauthor{\bsnm{Schacht}, \binits{A.L.}}:
\batitle{How to improve r\&d productivity: the pharmaceutical industry's grand
  challenge}.
\bjtitle{Nature reviews Drug discovery}
\bvolume{9}(\bissue{3}),
\bfpage{203}--\blpage{214}
(\byear{2010})
\end{barticle}
\endbibitem

\bibitem[\protect\citeauthoryear{DiMasi et~al.}{2016}]{dimasi2016innovation}
\begin{barticle}
\bauthor{\bsnm{DiMasi}, \binits{J.A.}},
\bauthor{\bsnm{Grabowski}, \binits{H.G.}},
\bauthor{\bsnm{Hansen}, \binits{R.W.}}:
\batitle{Innovation in the pharmaceutical industry: new estimates of r\&d
  costs}.
\bjtitle{Journal of health economics}
\bvolume{47},
\bfpage{20}--\blpage{33}
(\byear{2016})
\end{barticle}
\endbibitem

\bibitem[\protect\citeauthoryear{Wouters et~al.}{2020}]{wouters2020estimated}
\begin{barticle}
\bauthor{\bsnm{Wouters}, \binits{O.J.}},
\bauthor{\bsnm{McKee}, \binits{M.}},
\bauthor{\bsnm{Luyten}, \binits{J.}}:
\batitle{Estimated research and development investment needed to bring a new
  medicine to market, 2009-2018}.
\bjtitle{Jama}
\bvolume{323}(\bissue{9}),
\bfpage{844}--\blpage{853}
(\byear{2020})
\end{barticle}
\endbibitem

\bibitem[\protect\citeauthoryear{Southey and
  Brunavs}{2023}]{southey2023introduction}
\begin{barticle}
\bauthor{\bsnm{Southey}, \binits{M.W.}},
\bauthor{\bsnm{Brunavs}, \binits{M.}}:
\batitle{Introduction to small molecule drug discovery and preclinical
  development}.
\bjtitle{Frontiers in Drug Discovery}
\bvolume{3},
\bfpage{1314077}
(\byear{2023})
\end{barticle}
\endbibitem

\bibitem[\protect\citeauthoryear{Lu et~al.}{2020}]{lu2020development}
\begin{barticle}
\bauthor{\bsnm{Lu}, \binits{R.-M.}},
\bauthor{\bsnm{Hwang}, \binits{Y.-C.}},
\bauthor{\bsnm{Liu}, \binits{I.-J.}},
\bauthor{\bsnm{Lee}, \binits{C.-C.}},
\bauthor{\bsnm{Tsai}, \binits{H.-Z.}},
\bauthor{\bsnm{Li}, \binits{H.-J.}},
\bauthor{\bsnm{Wu}, \binits{H.-C.}}:
\batitle{Development of therapeutic antibodies for the treatment of diseases}.
\bjtitle{Journal of biomedical science}
\bvolume{27},
\bfpage{1}--\blpage{30}
(\byear{2020})
\end{barticle}
\endbibitem

\bibitem[\protect\citeauthoryear{Sadybekov and
  Katritch}{2023}]{sadybekov2023computational}
\begin{barticle}
\bauthor{\bsnm{Sadybekov}, \binits{A.V.}},
\bauthor{\bsnm{Katritch}, \binits{V.}}:
\batitle{Computational approaches streamlining drug discovery}.
\bjtitle{Nature}
\bvolume{616}(\bissue{7958}),
\bfpage{673}--\blpage{685}
(\byear{2023})
\end{barticle}
\endbibitem

\bibitem[\protect\citeauthoryear{Huang et~al.}{2024}]{huang2024ai}
\begin{barticle}
\bauthor{\bsnm{Huang}, \binits{D.}},
\bauthor{\bsnm{Yang}, \binits{M.}},
\bauthor{\bsnm{Wen}, \binits{X.}},
\bauthor{\bsnm{Xia}, \binits{S.}},
\bauthor{\bsnm{Yuan}, \binits{B.}}:
\batitle{Ai-driven drug discovery:: Accelerating the development of novel
  therapeutics in biopharmaceuticals}.
\bjtitle{Journal of Knowledge Learning and Science Technology ISSN: 2959-6386
  (online)}
\bvolume{3}(\bissue{3}),
\bfpage{206}--\blpage{224}
(\byear{2024})
\end{barticle}
\endbibitem

\bibitem[\protect\citeauthoryear{Son et~al.}{2024}]{son2024revolutionizing}
\begin{barticle}
\bauthor{\bsnm{Son}, \binits{A.}},
\bauthor{\bsnm{Park}, \binits{J.}},
\bauthor{\bsnm{Kim}, \binits{W.}},
\bauthor{\bsnm{Yoon}, \binits{Y.}},
\bauthor{\bsnm{Lee}, \binits{S.}},
\bauthor{\bsnm{Park}, \binits{Y.}},
\bauthor{\bsnm{Kim}, \binits{H.}}:
\batitle{Revolutionizing molecular design for innovative therapeutic
  applications through artificial intelligence}.
\bjtitle{Molecules}
\bvolume{29}(\bissue{19}),
\bfpage{4626}
(\byear{2024})
\end{barticle}
\endbibitem

\bibitem[\protect\citeauthoryear{Baslan and
  Hicks}{2017}]{baslan2017unravelling}
\begin{barticle}
\bauthor{\bsnm{Baslan}, \binits{T.}},
\bauthor{\bsnm{Hicks}, \binits{J.}}:
\batitle{Unravelling biology and shifting paradigms in cancer with single-cell
  sequencing}.
\bjtitle{Nature Reviews Cancer}
\bvolume{17}(\bissue{9}),
\bfpage{557}--\blpage{569}
(\byear{2017})
\end{barticle}
\endbibitem

\bibitem[\protect\citeauthoryear{Ofengeim et~al.}{2017}]{ofengeim2017single}
\begin{barticle}
\bauthor{\bsnm{Ofengeim}, \binits{D.}},
\bauthor{\bsnm{Giagtzoglou}, \binits{N.}},
\bauthor{\bsnm{Huh}, \binits{D.}},
\bauthor{\bsnm{Zou}, \binits{C.}},
\bauthor{\bsnm{Yuan}, \binits{J.}}:
\batitle{Single-cell rna sequencing: unraveling the brain one cell at a time}.
\bjtitle{Trends in molecular medicine}
\bvolume{23}(\bissue{6}),
\bfpage{563}--\blpage{576}
(\byear{2017})
\end{barticle}
\endbibitem

\bibitem[\protect\citeauthoryear{Rozenblatt-Rosen
  et~al.}{2017}]{rozenblatt2017human}
\begin{barticle}
\bauthor{\bsnm{Rozenblatt-Rosen}, \binits{O.}},
\bauthor{\bsnm{Stubbington}, \binits{M.J.}},
\bauthor{\bsnm{Regev}, \binits{A.}},
\bauthor{\bsnm{Teichmann}, \binits{S.A.}}:
\batitle{The human cell atlas: from vision to reality}.
\bjtitle{Nature}
\bvolume{550}(\bissue{7677}),
\bfpage{451}--\blpage{453}
(\byear{2017})
\end{barticle}
\endbibitem

\bibitem[\protect\citeauthoryear{Potter}{2018}]{potter2018single}
\begin{barticle}
\bauthor{\bsnm{Potter}, \binits{S.S.}}:
\batitle{Single-cell rna sequencing for the study of development, physiology
  and disease}.
\bjtitle{Nature Reviews Nephrology}
\bvolume{14}(\bissue{8}),
\bfpage{479}--\blpage{492}
(\byear{2018})
\end{barticle}
\endbibitem

\bibitem[\protect\citeauthoryear{Van~de Sande
  et~al.}{2023}]{van2023applications}
\begin{barticle}
\bauthor{\bsnm{Sande}, \binits{B.}},
\bauthor{\bsnm{Lee}, \binits{J.S.}},
\bauthor{\bsnm{Mutasa-Gottgens}, \binits{E.}},
\bauthor{\bsnm{Naughton}, \binits{B.}},
\bauthor{\bsnm{Bacon}, \binits{W.}},
\bauthor{\bsnm{Manning}, \binits{J.}},
\bauthor{\bsnm{Wang}, \binits{Y.}},
\bauthor{\bsnm{Pollard}, \binits{J.}},
\bauthor{\bsnm{Mendez}, \binits{M.}},
\bauthor{\bsnm{Hill}, \binits{J.}}, \betal:
\batitle{Applications of single-cell rna sequencing in drug discovery and
  development}.
\bjtitle{Nature Reviews Drug Discovery}
\bvolume{22}(\bissue{6}),
\bfpage{496}--\blpage{520}
(\byear{2023})
\end{barticle}
\endbibitem

\bibitem[\protect\citeauthoryear{Dann et~al.}{2024}]{Dann2024.04.04.24305313}
\begin{barticle}
\bauthor{\bsnm{Dann}, \binits{E.}},
\bauthor{\bsnm{Teeple}, \binits{E.}},
\bauthor{\bsnm{Elmentaite}, \binits{R.}},
\bauthor{\bsnm{Meyer}, \binits{K.B.}},
\bauthor{\bsnm{Gaglia}, \binits{G.}},
\bauthor{\bsnm{Nestle}, \binits{F.}},
\bauthor{\bsnm{Savova}, \binits{V.}},
\bauthor{\bsnm{Rinaldis}, \binits{E.}},
\bauthor{\bsnm{Teichmann}, \binits{S.A.}}:
\batitle{Estimating the impact of single-cell rna sequencing of human tissues
  on drug target validation}.
\bjtitle{medRxiv}
(\byear{2024})
\doiurl{10.1101/2024.04.04.24305313}
{\href{https://arxiv.org/abs/https://www.medrxiv.org/content/early/2024/10/22/2024.04.04.24305313.full.pdf}{{https://www.medrxiv.org/content/early/2024/10/22/2024.04.04.24305313.full.pdf}}}
\end{barticle}
\endbibitem

\bibitem[\protect\citeauthoryear{Tang et~al.}{2024}]{tang2024survey}
\begin{barticle}
\bauthor{\bsnm{Tang}, \binits{X.}},
\bauthor{\bsnm{Dai}, \binits{H.}},
\bauthor{\bsnm{Knight}, \binits{E.}},
\bauthor{\bsnm{Wu}, \binits{F.}},
\bauthor{\bsnm{Li}, \binits{Y.}},
\bauthor{\bsnm{Li}, \binits{T.}},
\bauthor{\bsnm{Gerstein}, \binits{M.}}:
\batitle{A survey of generative ai for de novo drug design: new frontiers in
  molecule and protein generation}.
\bjtitle{Briefings in Bioinformatics}
\bvolume{25}(\bissue{4}),
\bfpage{338}
(\byear{2024})
\end{barticle}
\endbibitem

\bibitem[\protect\citeauthoryear{Shanehsazzadeh
  et~al.}{2023}]{shanehsazzadeh2023unlocking}
\begin{botherref}
\oauthor{\bsnm{Shanehsazzadeh}, \binits{A.}},
\oauthor{\bsnm{Bachas}, \binits{S.}},
\oauthor{\bsnm{McPartlon}, \binits{M.}},
\oauthor{\bsnm{Kasun}, \binits{G.}},
\oauthor{\bsnm{Sutton}, \binits{J.M.}},
\oauthor{\bsnm{Steiger}, \binits{A.K.}},
\oauthor{\bsnm{Shuai}, \binits{R.}},
\oauthor{\bsnm{Kohnert}, \binits{C.}},
\oauthor{\bsnm{Rakocevic}, \binits{G.}},
\oauthor{\bsnm{Gutierrez}, \binits{J.M.}}, et al.:
Unlocking de novo antibody design with generative artificial intelligence.
bioRxiv,
2023--01
(2023)
\end{botherref}
\endbibitem

\bibitem[\protect\citeauthoryear{Swanson et~al.}{2024}]{swanson2024generative}
\begin{barticle}
\bauthor{\bsnm{Swanson}, \binits{K.}},
\bauthor{\bsnm{Liu}, \binits{G.}},
\bauthor{\bsnm{Catacutan}, \binits{D.B.}},
\bauthor{\bsnm{Arnold}, \binits{A.}},
\bauthor{\bsnm{Zou}, \binits{J.}},
\bauthor{\bsnm{Stokes}, \binits{J.M.}}:
\batitle{Generative ai for designing and validating easily synthesizable and
  structurally novel antibiotics}.
\bjtitle{Nature Machine Intelligence}
\bvolume{6}(\bissue{3}),
\bfpage{338}--\blpage{353}
(\byear{2024})
\end{barticle}
\endbibitem

\bibitem[\protect\citeauthoryear{Athaya et~al.}{2023}]{athaya2023multimodal}
\begin{barticle}
\bauthor{\bsnm{Athaya}, \binits{T.}},
\bauthor{\bsnm{Ripan}, \binits{R.C.}},
\bauthor{\bsnm{Li}, \binits{X.}},
\bauthor{\bsnm{Hu}, \binits{H.}}:
\batitle{Multimodal deep learning approaches for single-cell multi-omics data
  integration}.
\bjtitle{Briefings in Bioinformatics}
\bvolume{24}(\bissue{5}),
\bfpage{313}
(\byear{2023})
\end{barticle}
\endbibitem

\bibitem[\protect\citeauthoryear{Sza{\l}ata
  et~al.}{2024}]{szalata2024transformers}
\begin{barticle}
\bauthor{\bsnm{Sza{\l}ata}, \binits{A.}},
\bauthor{\bsnm{Hrovatin}, \binits{K.}},
\bauthor{\bsnm{Becker}, \binits{S.}},
\bauthor{\bsnm{Tejada-Lapuerta}, \binits{A.}},
\bauthor{\bsnm{Cui}, \binits{H.}},
\bauthor{\bsnm{Wang}, \binits{B.}},
\bauthor{\bsnm{Theis}, \binits{F.J.}}:
\batitle{Transformers in single-cell omics: a review and new perspectives}.
\bjtitle{Nature methods}
\bvolume{21}(\bissue{8}),
\bfpage{1430}--\blpage{1443}
(\byear{2024})
\end{barticle}
\endbibitem

\bibitem[\protect\citeauthoryear{Jumper et~al.}{2021}]{alphafold}
\begin{barticle}
\bauthor{\bsnm{Jumper}, \binits{J.}},
\bauthor{\bsnm{Evans}, \binits{R.}},
\bauthor{\bsnm{Pritzel}, \binits{A.}},
\bauthor{\bsnm{Green}, \binits{T.}},
\bauthor{\bsnm{Figurnov}, \binits{M.}},
\bauthor{\bsnm{Ronneberger}, \binits{O.}},
\bauthor{\bsnm{Tunyasuvunakool}, \binits{K.}},
\bauthor{\bsnm{Bates}, \binits{R.}},
\bauthor{\bsnm{{\v Z}{\'\i}dek}, \binits{A.}},
\bauthor{\bsnm{Potapenko}, \binits{A.}},
\bauthor{\bsnm{Bridgland}, \binits{A.}},
\bauthor{\bsnm{Meyer}, \binits{C.}},
\bauthor{\bsnm{Kohl}, \binits{S.A.A.}},
\bauthor{\bsnm{Ballard}, \binits{A.J.}},
\bauthor{\bsnm{Cowie}, \binits{A.}},
\bauthor{\bsnm{Romera-Paredes}, \binits{B.}},
\bauthor{\bsnm{Nikolov}, \binits{S.}},
\bauthor{\bsnm{Jain}, \binits{R.}},
\bauthor{\bsnm{Adler}, \binits{J.}},
\bauthor{\bsnm{Back}, \binits{T.}},
\bauthor{\bsnm{Petersen}, \binits{S.}},
\bauthor{\bsnm{Reiman}, \binits{D.}},
\bauthor{\bsnm{Clancy}, \binits{E.}},
\bauthor{\bsnm{Zielinski}, \binits{M.}},
\bauthor{\bsnm{Steinegger}, \binits{M.}},
\bauthor{\bsnm{Pacholska}, \binits{M.}},
\bauthor{\bsnm{Berghammer}, \binits{T.}},
\bauthor{\bsnm{Bodenstein}, \binits{S.}},
\bauthor{\bsnm{Silver}, \binits{D.}},
\bauthor{\bsnm{Vinyals}, \binits{O.}},
\bauthor{\bsnm{Senior}, \binits{A.W.}},
\bauthor{\bsnm{Kavukcuoglu}, \binits{K.}},
\bauthor{\bsnm{Kohli}, \binits{P.}},
\bauthor{\bsnm{Hassabis}, \binits{D.}}:
\batitle{Highly accurate protein structure prediction with {AlphaFold}}.
\bjtitle{Nature}
\bvolume{596}(\bissue{7873}),
\bfpage{583}--\blpage{589}
(\byear{2021})
\end{barticle}
\endbibitem

\bibitem[\protect\citeauthoryear{Evans et~al.}{2022}]{alphafold-multimer}
\begin{barticle}
\bauthor{\bsnm{Evans}, \binits{R.}},
\bauthor{\bsnm{O{\textquoteright}Neill}, \binits{M.}},
\bauthor{\bsnm{Pritzel}, \binits{A.}},
\bauthor{\bsnm{Antropova}, \binits{N.}},
\bauthor{\bsnm{Senior}, \binits{A.}},
\bauthor{\bsnm{Green}, \binits{T.}},
\bauthor{\bsnm{{\v Z}{\'\i}dek}, \binits{A.}},
\bauthor{\bsnm{Bates}, \binits{R.}},
\bauthor{\bsnm{Blackwell}, \binits{S.}},
\bauthor{\bsnm{Yim}, \binits{J.}},
\bauthor{\bsnm{Ronneberger}, \binits{O.}},
\bauthor{\bsnm{Bodenstein}, \binits{S.}},
\bauthor{\bsnm{Zielinski}, \binits{M.}},
\bauthor{\bsnm{Bridgland}, \binits{A.}},
\bauthor{\bsnm{Potapenko}, \binits{A.}},
\bauthor{\bsnm{Cowie}, \binits{A.}},
\bauthor{\bsnm{Tunyasuvunakool}, \binits{K.}},
\bauthor{\bsnm{Jain}, \binits{R.}},
\bauthor{\bsnm{Clancy}, \binits{E.}},
\bauthor{\bsnm{Kohli}, \binits{P.}},
\bauthor{\bsnm{Jumper}, \binits{J.}},
\bauthor{\bsnm{Hassabis}, \binits{D.}}:
\batitle{Protein complex prediction with alphafold-multimer}.
\bjtitle{bioRxiv}
(\byear{2022})
\doiurl{10.1101/2021.10.04.463034}
{\href{https://arxiv.org/abs/https://www.biorxiv.org/content/early/2022/03/10/2021.10.04.463034.full.pdf}{{https://www.biorxiv.org/content/early/2022/03/10/2021.10.04.463034.full.pdf}}}
\end{barticle}
\endbibitem

\bibitem[\protect\citeauthoryear{Abramson et~al.}{2024}]{alphafold3}
\begin{barticle}
\bauthor{\bsnm{Abramson}, \binits{J.}},
\bauthor{\bsnm{Adler}, \binits{J.}},
\bauthor{\bsnm{Dunger}, \binits{J.}},
\bauthor{\bsnm{Evans}, \binits{R.}},
\bauthor{\bsnm{Green}, \binits{T.}},
\bauthor{\bsnm{Pritzel}, \binits{A.}},
\bauthor{\bsnm{Ronneberger}, \binits{O.}},
\bauthor{\bsnm{Willmore}, \binits{L.}},
\bauthor{\bsnm{Ballard}, \binits{A.J.}},
\bauthor{\bsnm{Bambrick}, \binits{J.}},
\bauthor{\bsnm{Bodenstein}, \binits{S.W.}},
\bauthor{\bsnm{Evans}, \binits{D.A.}},
\bauthor{\bsnm{Hung}, \binits{C.-C.}},
\bauthor{\bsnm{O'Neill}, \binits{M.}},
\bauthor{\bsnm{Reiman}, \binits{D.}},
\bauthor{\bsnm{Tunyasuvunakool}, \binits{K.}},
\bauthor{\bsnm{Wu}, \binits{Z.}},
\bauthor{\bsnm{{\v Z}emgulyt{\.e}}, \binits{A.}},
\bauthor{\bsnm{Arvaniti}, \binits{E.}},
\bauthor{\bsnm{Beattie}, \binits{C.}},
\bauthor{\bsnm{Bertolli}, \binits{O.}},
\bauthor{\bsnm{Bridgland}, \binits{A.}},
\bauthor{\bsnm{Cherepanov}, \binits{A.}},
\bauthor{\bsnm{Congreve}, \binits{M.}},
\bauthor{\bsnm{Cowen-Rivers}, \binits{A.I.}},
\bauthor{\bsnm{Cowie}, \binits{A.}},
\bauthor{\bsnm{Figurnov}, \binits{M.}},
\bauthor{\bsnm{Fuchs}, \binits{F.B.}},
\bauthor{\bsnm{Gladman}, \binits{H.}},
\bauthor{\bsnm{Jain}, \binits{R.}},
\bauthor{\bsnm{Khan}, \binits{Y.A.}},
\bauthor{\bsnm{Low}, \binits{C.M.R.}},
\bauthor{\bsnm{Perlin}, \binits{K.}},
\bauthor{\bsnm{Potapenko}, \binits{A.}},
\bauthor{\bsnm{Savy}, \binits{P.}},
\bauthor{\bsnm{Singh}, \binits{S.}},
\bauthor{\bsnm{Stecula}, \binits{A.}},
\bauthor{\bsnm{Thillaisundaram}, \binits{A.}},
\bauthor{\bsnm{Tong}, \binits{C.}},
\bauthor{\bsnm{Yakneen}, \binits{S.}},
\bauthor{\bsnm{Zhong}, \binits{E.D.}},
\bauthor{\bsnm{Zielinski}, \binits{M.}},
\bauthor{\bsnm{{\v Z}{\'\i}dek}, \binits{A.}},
\bauthor{\bsnm{Bapst}, \binits{V.}},
\bauthor{\bsnm{Kohli}, \binits{P.}},
\bauthor{\bsnm{Jaderberg}, \binits{M.}},
\bauthor{\bsnm{Hassabis}, \binits{D.}},
\bauthor{\bsnm{Jumper}, \binits{J.M.}}:
\batitle{Accurate structure prediction of biomolecular interactions with
  {AlphaFold} 3}.
\bjtitle{Nature}
\bvolume{630}(\bissue{8016}),
\bfpage{493}--\blpage{500}
(\byear{2024})
\end{barticle}
\endbibitem

\bibitem[\protect\citeauthoryear{Xu et~al.}{2023}]{xu2023ciform}
\begin{barticle}
\bauthor{\bsnm{Xu}, \binits{J.}},
\bauthor{\bsnm{Zhang}, \binits{A.}},
\bauthor{\bsnm{Liu}, \binits{F.}},
\bauthor{\bsnm{Chen}, \binits{L.}},
\bauthor{\bsnm{Zhang}, \binits{X.}}:
\batitle{Ciform as a transformer-based model for cell-type annotation of
  large-scale single-cell rna-seq data}.
\bjtitle{Briefings in Bioinformatics}
\bvolume{24}(\bissue{4}),
\bfpage{195}
(\byear{2023})
\end{barticle}
\endbibitem

\bibitem[\protect\citeauthoryear{Ross et~al.}{2022}]{ross2022large}
\begin{barticle}
\bauthor{\bsnm{Ross}, \binits{J.}},
\bauthor{\bsnm{Belgodere}, \binits{B.}},
\bauthor{\bsnm{Chenthamarakshan}, \binits{V.}},
\bauthor{\bsnm{Padhi}, \binits{I.}},
\bauthor{\bsnm{Mroueh}, \binits{Y.}},
\bauthor{\bsnm{Das}, \binits{P.}}:
\batitle{Large-scale chemical language representations capture molecular
  structure and properties}.
\bjtitle{Nature Machine Intelligence}
\bvolume{4}(\bissue{12}),
\bfpage{1256}--\blpage{1264}
(\byear{2022})
\end{barticle}
\endbibitem

\bibitem[\protect\citeauthoryear{Chaves et~al.}{2024}]{chaves2024tx}
\begin{botherref}
\oauthor{\bsnm{Chaves}, \binits{J.M.Z.}},
\oauthor{\bsnm{Wang}, \binits{E.}},
\oauthor{\bsnm{Tu}, \binits{T.}},
\oauthor{\bsnm{Vaishnav}, \binits{E.D.}},
\oauthor{\bsnm{Lee}, \binits{B.}},
\oauthor{\bsnm{Mahdavi}, \binits{S.S.}},
\oauthor{\bsnm{Semturs}, \binits{C.}},
\oauthor{\bsnm{Fleet}, \binits{D.}},
\oauthor{\bsnm{Natarajan}, \binits{V.}},
\oauthor{\bsnm{Azizi}, \binits{S.}}:
Tx-llm: A large language model for therapeutics.
arXiv preprint arXiv:2406.06316
(2024)
\end{botherref}
\endbibitem

\bibitem[\protect\citeauthoryear{Liu et~al.}{2020}]{liu2020deepcdr}
\begin{barticle}
\bauthor{\bsnm{Liu}, \binits{Q.}},
\bauthor{\bsnm{Hu}, \binits{Z.}},
\bauthor{\bsnm{Jiang}, \binits{R.}},
\bauthor{\bsnm{Zhou}, \binits{M.}}:
\batitle{Deepcdr: a hybrid graph convolutional network for predicting cancer
  drug response}.
\bjtitle{Bioinformatics}
\bvolume{36}(\bissue{Supplement\_2}),
\bfpage{911}--\blpage{918}
(\byear{2020})
\end{barticle}
\endbibitem

\bibitem[\protect\citeauthoryear{Kong et~al.}{2023}]{kong2023end}
\begin{botherref}
\oauthor{\bsnm{Kong}, \binits{X.}},
\oauthor{\bsnm{Huang}, \binits{W.}},
\oauthor{\bsnm{Liu}, \binits{Y.}}:
End-to-end full-atom antibody design.
arXiv preprint arXiv:2302.00203
(2023)
\end{botherref}
\endbibitem

\bibitem[\protect\citeauthoryear{Jing et~al.}{2024}]{jing2024accurate}
\begin{barticle}
\bauthor{\bsnm{Jing}, \binits{H.}},
\bauthor{\bsnm{Gao}, \binits{Z.}},
\bauthor{\bsnm{Xu}, \binits{S.}},
\bauthor{\bsnm{Shen}, \binits{T.}},
\bauthor{\bsnm{Peng}, \binits{Z.}},
\bauthor{\bsnm{He}, \binits{S.}},
\bauthor{\bsnm{You}, \binits{T.}},
\bauthor{\bsnm{Ye}, \binits{S.}},
\bauthor{\bsnm{Lin}, \binits{W.}},
\bauthor{\bsnm{Sun}, \binits{S.}}:
\batitle{Accurate prediction of antibody function and structure using
  bio-inspired antibody language model}.
\bjtitle{Briefings in Bioinformatics}
\bvolume{25}(\bissue{4}),
\bfpage{245}
(\byear{2024})
\end{barticle}
\endbibitem

\bibitem[\protect\citeauthoryear{Weber et~al.}{2021}]{weber2021titan}
\begin{barticle}
\bauthor{\bsnm{Weber}, \binits{A.}},
\bauthor{\bsnm{Born}, \binits{J.}},
\bauthor{\bsnm{Rodriguez~Mart{\'\i}nez}, \binits{M.}}:
\batitle{Titan: T-cell receptor specificity prediction with bimodal attention
  networks}.
\bjtitle{Bioinformatics}
\bvolume{37}(\bissue{Supplement\_1}),
\bfpage{237}--\blpage{244}
(\byear{2021})
\end{barticle}
\endbibitem

\bibitem[\protect\citeauthoryear{Jin et~al.}{2024}]{jin2024attabseq}
\begin{barticle}
\bauthor{\bsnm{Jin}, \binits{R.}},
\bauthor{\bsnm{Ye}, \binits{Q.}},
\bauthor{\bsnm{Wang}, \binits{J.}},
\bauthor{\bsnm{Cao}, \binits{Z.}},
\bauthor{\bsnm{Jiang}, \binits{D.}},
\bauthor{\bsnm{Wang}, \binits{T.}},
\bauthor{\bsnm{Kang}, \binits{Y.}},
\bauthor{\bsnm{Xu}, \binits{W.}},
\bauthor{\bsnm{Hsieh}, \binits{C.-Y.}},
\bauthor{\bsnm{Hou}, \binits{T.}}:
\batitle{Attabseq: an attention-based deep learning prediction method for
  antigen--antibody binding affinity changes based on protein sequences}.
\bjtitle{Briefings in Bioinformatics}
\bvolume{25}(\bissue{4}),
\bfpage{304}
(\byear{2024})
\end{barticle}
\endbibitem

\bibitem[\protect\citeauthoryear{Xu et~al.}{2022}]{xu2022peer}
\begin{barticle}
\bauthor{\bsnm{Xu}, \binits{M.}},
\bauthor{\bsnm{Zhang}, \binits{Z.}},
\bauthor{\bsnm{Lu}, \binits{J.}},
\bauthor{\bsnm{Zhu}, \binits{Z.}},
\bauthor{\bsnm{Zhang}, \binits{Y.}},
\bauthor{\bsnm{Chang}, \binits{M.}},
\bauthor{\bsnm{Liu}, \binits{R.}},
\bauthor{\bsnm{Tang}, \binits{J.}}:
\batitle{Peer: a comprehensive and multi-task benchmark for protein sequence
  understanding}.
\bjtitle{Advances in Neural Information Processing Systems}
\bvolume{35},
\bfpage{35156}--\blpage{35173}
(\byear{2022})
\end{barticle}
\endbibitem

\bibitem[\protect\citeauthoryear{Qi et~al.}{2020}]{qi2020clustering}
\begin{barticle}
\bauthor{\bsnm{Qi}, \binits{R.}},
\bauthor{\bsnm{Ma}, \binits{A.}},
\bauthor{\bsnm{Ma}, \binits{Q.}},
\bauthor{\bsnm{Zou}, \binits{Q.}}:
\batitle{Clustering and classification methods for single-cell rna-sequencing
  data}.
\bjtitle{Briefings in bioinformatics}
\bvolume{21}(\bissue{4}),
\bfpage{1196}--\blpage{1208}
(\byear{2020})
\end{barticle}
\endbibitem

\bibitem[\protect\citeauthoryear{Cui et~al.}{2024}]{cui2024scgpt}
\begin{botherref}
\oauthor{\bsnm{Cui}, \binits{H.}},
\oauthor{\bsnm{Wang}, \binits{C.}},
\oauthor{\bsnm{Maan}, \binits{H.}},
\oauthor{\bsnm{Pang}, \binits{K.}},
\oauthor{\bsnm{Luo}, \binits{F.}},
\oauthor{\bsnm{Duan}, \binits{N.}},
\oauthor{\bsnm{Wang}, \binits{B.}}:
scgpt: toward building a foundation model for single-cell multi-omics using
  generative ai.
Nature Methods,
1--11
(2024)
\end{botherref}
\endbibitem

\bibitem[\protect\citeauthoryear{Yang et~al.}{2022}]{yang2022scbert}
\begin{barticle}
\bauthor{\bsnm{Yang}, \binits{F.}},
\bauthor{\bsnm{Wang}, \binits{W.}},
\bauthor{\bsnm{Wang}, \binits{F.}},
\bauthor{\bsnm{Fang}, \binits{Y.}},
\bauthor{\bsnm{Tang}, \binits{D.}},
\bauthor{\bsnm{Huang}, \binits{J.}},
\bauthor{\bsnm{Lu}, \binits{H.}},
\bauthor{\bsnm{Yao}, \binits{J.}}:
\batitle{scbert as a large-scale pretrained deep language model for cell type
  annotation of single-cell rna-seq data}.
\bjtitle{Nature Machine Intelligence}
\bvolume{4}(\bissue{10}),
\bfpage{852}--\blpage{866}
(\byear{2022})
\end{barticle}
\endbibitem

\bibitem[\protect\citeauthoryear{Zheng et~al.}{2017}]{zheng2017massively}
\begin{barticle}
\bauthor{\bsnm{Zheng}, \binits{G.X.}},
\bauthor{\bsnm{Terry}, \binits{J.M.}},
\bauthor{\bsnm{Belgrader}, \binits{P.}},
\bauthor{\bsnm{Ryvkin}, \binits{P.}},
\bauthor{\bsnm{Bent}, \binits{Z.W.}},
\bauthor{\bsnm{Wilson}, \binits{R.}},
\bauthor{\bsnm{Ziraldo}, \binits{S.B.}},
\bauthor{\bsnm{Wheeler}, \binits{T.D.}},
\bauthor{\bsnm{McDermott}, \binits{G.P.}},
\bauthor{\bsnm{Zhu}, \binits{J.}}, \betal:
\batitle{Massively parallel digital transcriptional profiling of single cells}.
\bjtitle{Nature communications}
\bvolume{8}(\bissue{1}),
\bfpage{14049}
(\byear{2017})
\end{barticle}
\endbibitem

\bibitem[\protect\citeauthoryear{Theodoris et~al.}{2023}]{genformer}
\begin{barticle}
\bauthor{\bsnm{Theodoris}, \binits{C.V.}},
\bauthor{\bsnm{Xiao}, \binits{L.}},
\bauthor{\bsnm{Chopra}, \binits{A.}},
\bauthor{\bsnm{Chaffin}, \binits{M.D.}},
\bauthor{\bsnm{Al~Sayed}, \binits{Z.R.}},
\bauthor{\bsnm{Hill}, \binits{M.C.}},
\bauthor{\bsnm{Mantineo}, \binits{H.}},
\bauthor{\bsnm{Brydon}, \binits{E.M.}},
\bauthor{\bsnm{Zeng}, \binits{Z.}},
\bauthor{\bsnm{Liu}, \binits{X.S.}},
\bauthor{\bsnm{Ellinor}, \binits{P.T.}}:
\batitle{Transfer learning enables predictions in network biology}.
\bjtitle{Nature}
\bvolume{618}(\bissue{7965}),
\bfpage{616}--\blpage{624}
(\byear{2023})
\doiurl{10.1038/s41586-023-06139-9}
\end{barticle}
\endbibitem

\bibitem[\protect\citeauthoryear{Wu et~al.}{2018}]{wu2018moleculenet}
\begin{barticle}
\bauthor{\bsnm{Wu}, \binits{Z.}},
\bauthor{\bsnm{Ramsundar}, \binits{B.}},
\bauthor{\bsnm{Feinberg}, \binits{E.N.}},
\bauthor{\bsnm{Gomes}, \binits{J.}},
\bauthor{\bsnm{Geniesse}, \binits{C.}},
\bauthor{\bsnm{Pappu}, \binits{A.S.}},
\bauthor{\bsnm{Leswing}, \binits{K.}},
\bauthor{\bsnm{Pande}, \binits{V.}}:
\batitle{Moleculenet: a benchmark for molecular machine learning}.
\bjtitle{Chemical science}
\bvolume{9}(\bissue{2}),
\bfpage{513}--\blpage{530}
(\byear{2018})
\end{barticle}
\endbibitem

\bibitem[\protect\citeauthoryear{Barretina et~al.}{2012}]{barretina2012cancer}
\begin{barticle}
\bauthor{\bsnm{Barretina}, \binits{J.}},
\bauthor{\bsnm{Caponigro}, \binits{G.}},
\bauthor{\bsnm{Stransky}, \binits{N.}},
\bauthor{\bsnm{Venkatesan}, \binits{K.}},
\bauthor{\bsnm{Margolin}, \binits{A.A.}},
\bauthor{\bsnm{Kim}, \binits{S.}},
\bauthor{\bsnm{Wilson}, \binits{C.J.}},
\bauthor{\bsnm{Leh{\'a}r}, \binits{J.}},
\bauthor{\bsnm{Kryukov}, \binits{G.V.}},
\bauthor{\bsnm{Sonkin}, \binits{D.}}, \betal:
\batitle{The cancer cell line encyclopedia enables predictive modelling of
  anticancer drug sensitivity}.
\bjtitle{Nature}
\bvolume{483}(\bissue{7391}),
\bfpage{603}--\blpage{607}
(\byear{2012})
\end{barticle}
\endbibitem

\bibitem[\protect\citeauthoryear{Yang et~al.}{2012}]{yang2012genomics}
\begin{barticle}
\bauthor{\bsnm{Yang}, \binits{W.}},
\bauthor{\bsnm{Soares}, \binits{J.}},
\bauthor{\bsnm{Greninger}, \binits{P.}},
\bauthor{\bsnm{Edelman}, \binits{E.J.}},
\bauthor{\bsnm{Lightfoot}, \binits{H.}},
\bauthor{\bsnm{Forbes}, \binits{S.}},
\bauthor{\bsnm{Bindal}, \binits{N.}},
\bauthor{\bsnm{Beare}, \binits{D.}},
\bauthor{\bsnm{Smith}, \binits{J.A.}},
\bauthor{\bsnm{Thompson}, \binits{I.R.}}, \betal:
\batitle{Genomics of drug sensitivity in cancer (gdsc): a resource for
  therapeutic biomarker discovery in cancer cells}.
\bjtitle{Nucleic acids research}
\bvolume{41}(\bissue{D1}),
\bfpage{955}--\blpage{961}
(\byear{2012})
\end{barticle}
\endbibitem

\bibitem[\protect\citeauthoryear{Lind and Anderson}{2019}]{lind2019predicting}
\begin{barticle}
\bauthor{\bsnm{Lind}, \binits{A.P.}},
\bauthor{\bsnm{Anderson}, \binits{P.C.}}:
\batitle{Predicting drug activity against cancer cells by random forest models
  based on minimal genomic information and chemical properties}.
\bjtitle{PloS one}
\bvolume{14}(\bissue{7}),
\bfpage{0219774}
(\byear{2019})
\end{barticle}
\endbibitem

\bibitem[\protect\citeauthoryear{Liu et~al.}{2022}]{liu2022graphcdr}
\begin{barticle}
\bauthor{\bsnm{Liu}, \binits{X.}},
\bauthor{\bsnm{Song}, \binits{C.}},
\bauthor{\bsnm{Huang}, \binits{F.}},
\bauthor{\bsnm{Fu}, \binits{H.}},
\bauthor{\bsnm{Xiao}, \binits{W.}},
\bauthor{\bsnm{Zhang}, \binits{W.}}:
\batitle{Graphcdr: a graph neural network method with contrastive learning for
  cancer drug response prediction}.
\bjtitle{Briefings in Bioinformatics}
\bvolume{23}(\bissue{1}),
\bfpage{457}
(\byear{2022})
\end{barticle}
\endbibitem

\bibitem[\protect\citeauthoryear{Huang et~al.}{2021}]{huang2021therapeutics}
\begin{botherref}
\oauthor{\bsnm{Huang}, \binits{K.}},
\oauthor{\bsnm{Fu}, \binits{T.}},
\oauthor{\bsnm{Gao}, \binits{W.}},
\oauthor{\bsnm{Zhao}, \binits{Y.}},
\oauthor{\bsnm{Roohani}, \binits{Y.}},
\oauthor{\bsnm{Leskovec}, \binits{J.}},
\oauthor{\bsnm{Coley}, \binits{C.W.}},
\oauthor{\bsnm{Xiao}, \binits{C.}},
\oauthor{\bsnm{Sun}, \binits{J.}},
\oauthor{\bsnm{Zitnik}, \binits{M.}}:
Therapeutics data commons: Machine learning datasets and tasks for drug
  discovery and development.
arXiv preprint arXiv:2102.09548
(2021)
\end{botherref}
\endbibitem

\bibitem[\protect\citeauthoryear{Weinstein et~al.}{2013}]{weinstein2013cancer}
\begin{barticle}
\bauthor{\bsnm{Weinstein}, \binits{J.N.}},
\bauthor{\bsnm{Collisson}, \binits{E.A.}},
\bauthor{\bsnm{Mills}, \binits{G.B.}},
\bauthor{\bsnm{Shaw}, \binits{K.R.}},
\bauthor{\bsnm{Ozenberger}, \binits{B.A.}},
\bauthor{\bsnm{Ellrott}, \binits{K.}},
\bauthor{\bsnm{Shmulevich}, \binits{I.}},
\bauthor{\bsnm{Sander}, \binits{C.}},
\bauthor{\bsnm{Stuart}, \binits{J.M.}}:
\batitle{The cancer genome atlas pan-cancer analysis project}.
\bjtitle{Nature genetics}
\bvolume{45}(\bissue{10}),
\bfpage{1113}--\blpage{1120}
(\byear{2013})
\end{barticle}
\endbibitem

\bibitem[\protect\citeauthoryear{Hummer et~al.}{2022}]{hummer2022advances}
\begin{barticle}
\bauthor{\bsnm{Hummer}, \binits{A.M.}},
\bauthor{\bsnm{Abanades}, \binits{B.}},
\bauthor{\bsnm{Deane}, \binits{C.M.}}:
\batitle{Advances in computational structure-based antibody design}.
\bjtitle{Current opinion in structural biology}
\bvolume{74},
\bfpage{102379}
(\byear{2022})
\end{barticle}
\endbibitem

\bibitem[\protect\citeauthoryear{Chiu et~al.}{2019}]{chiu2019antibody}
\begin{botherref}
\oauthor{\bsnm{Chiu}, \binits{M.}},
\oauthor{\bsnm{Goulet}, \binits{D.}},
\oauthor{\bsnm{Teplyakov}, \binits{A.}},
\oauthor{\bsnm{Gilliland}, \binits{G.}}:
Antibody structure and function: the basis for engineering therapeutics.
  Antibodies (Basel) 8 (4)
(2019)
\end{botherref}
\endbibitem

\bibitem[\protect\citeauthoryear{Basu et~al.}{2019}]{basu2019recombinant}
\begin{barticle}
\bauthor{\bsnm{Basu}, \binits{K.}},
\bauthor{\bsnm{Green}, \binits{E.M.}},
\bauthor{\bsnm{Cheng}, \binits{Y.}},
\bauthor{\bsnm{Craik}, \binits{C.S.}}:
\batitle{Why recombinant antibodies - benefits and applications}.
\bjtitle{Current opinion in biotechnology}
\bvolume{60},
\bfpage{153}--\blpage{158}
(\byear{2019})
\end{barticle}
\endbibitem

\bibitem[\protect\citeauthoryear{Carter and Lazar}{2018}]{carter2018next}
\begin{barticle}
\bauthor{\bsnm{Carter}, \binits{P.J.}},
\bauthor{\bsnm{Lazar}, \binits{G.A.}}:
\batitle{Next generation antibody drugs: pursuit of the'high-hanging fruit'}.
\bjtitle{Nature Reviews Drug Discovery}
\bvolume{17}(\bissue{3}),
\bfpage{197}--\blpage{223}
(\byear{2018})
\end{barticle}
\endbibitem

\bibitem[\protect\citeauthoryear{Beck et~al.}{2017}]{beck2017strategies}
\begin{barticle}
\bauthor{\bsnm{Beck}, \binits{A.}},
\bauthor{\bsnm{Goetsch}, \binits{L.}},
\bauthor{\bsnm{Dumontet}, \binits{C.}},
\bauthor{\bsnm{Corva{\"\i}a}, \binits{N.}}:
\batitle{Strategies and challenges for the next generation of antibody--drug
  conjugates}.
\bjtitle{Nature reviews Drug discovery}
\bvolume{16}(\bissue{5}),
\bfpage{315}--\blpage{337}
(\byear{2017})
\end{barticle}
\endbibitem

\bibitem[\protect\citeauthoryear{Saka et~al.}{2021}]{saka2021antibody}
\begin{barticle}
\bauthor{\bsnm{Saka}, \binits{K.}},
\bauthor{\bsnm{Kakuzaki}, \binits{T.}},
\bauthor{\bsnm{Metsugi}, \binits{S.}},
\bauthor{\bsnm{Kashiwagi}, \binits{D.}},
\bauthor{\bsnm{Yoshida}, \binits{K.}},
\bauthor{\bsnm{Wada}, \binits{M.}},
\bauthor{\bsnm{Tsunoda}, \binits{H.}},
\bauthor{\bsnm{Teramoto}, \binits{R.}}:
\batitle{Antibody design using lstm based deep generative model from phage
  display library for affinity maturation}.
\bjtitle{Scientific reports}
\bvolume{11}(\bissue{1}),
\bfpage{5852}
(\byear{2021})
\end{barticle}
\endbibitem

\bibitem[\protect\citeauthoryear{Jin et~al.}{2021}]{jin2021iterative}
\begin{botherref}
\oauthor{\bsnm{Jin}, \binits{W.}},
\oauthor{\bsnm{Wohlwend}, \binits{J.}},
\oauthor{\bsnm{Barzilay}, \binits{R.}},
\oauthor{\bsnm{Jaakkola}, \binits{T.}}:
Iterative refinement graph neural network for antibody sequence-structure
  co-design.
arXiv preprint arXiv:2110.04624
(2021)
\end{botherref}
\endbibitem

\bibitem[\protect\citeauthoryear{Jin et~al.}{2022}]{jin2022antibody}
\begin{botherref}
\oauthor{\bsnm{Jin}, \binits{W.}},
\oauthor{\bsnm{Barzilay}, \binits{R.}},
\oauthor{\bsnm{Jaakkola}, \binits{T.}}:
Antibody-antigen docking and design via hierarchical equivariant refinement.
arXiv preprint arXiv:2207.06616
(2022)
\end{botherref}
\endbibitem

\bibitem[\protect\citeauthoryear{Luo et~al.}{2022}]{luo2022antigen}
\begin{barticle}
\bauthor{\bsnm{Luo}, \binits{S.}},
\bauthor{\bsnm{Su}, \binits{Y.}},
\bauthor{\bsnm{Peng}, \binits{X.}},
\bauthor{\bsnm{Wang}, \binits{S.}},
\bauthor{\bsnm{Peng}, \binits{J.}},
\bauthor{\bsnm{Ma}, \binits{J.}}:
\batitle{Antigen-specific antibody design and optimization with diffusion-based
  generative models for protein structures}.
\bjtitle{Advances in Neural Information Processing Systems}
\bvolume{35},
\bfpage{9754}--\blpage{9767}
(\byear{2022})
\end{barticle}
\endbibitem

\bibitem[\protect\citeauthoryear{Kong et~al.}{2022}]{kong2022conditional}
\begin{botherref}
\oauthor{\bsnm{Kong}, \binits{X.}},
\oauthor{\bsnm{Huang}, \binits{W.}},
\oauthor{\bsnm{Liu}, \binits{Y.}}:
Conditional antibody design as 3d equivariant graph translation.
arXiv preprint arXiv:2208.06073
(2022)
\end{botherref}
\endbibitem

\bibitem[\protect\citeauthoryear{Zhou et~al.}{2024}]{zhou2024antigen}
\begin{botherref}
\oauthor{\bsnm{Zhou}, \binits{X.}},
\oauthor{\bsnm{Xue}, \binits{D.}},
\oauthor{\bsnm{Chen}, \binits{R.}},
\oauthor{\bsnm{Zheng}, \binits{Z.}},
\oauthor{\bsnm{Wang}, \binits{L.}},
\oauthor{\bsnm{Gu}, \binits{Q.}}:
Antigen-specific antibody design via direct energy-based preference
  optimization.
arXiv preprint arXiv:2403.16576
(2024)
\end{botherref}
\endbibitem

\bibitem[\protect\citeauthoryear{Dunbar et~al.}{2014}]{dunbar2014sabdab}
\begin{barticle}
\bauthor{\bsnm{Dunbar}, \binits{J.}},
\bauthor{\bsnm{Krawczyk}, \binits{K.}},
\bauthor{\bsnm{Leem}, \binits{J.}},
\bauthor{\bsnm{Baker}, \binits{T.}},
\bauthor{\bsnm{Fuchs}, \binits{A.}},
\bauthor{\bsnm{Georges}, \binits{G.}},
\bauthor{\bsnm{Shi}, \binits{J.}},
\bauthor{\bsnm{Deane}, \binits{C.M.}}:
\batitle{Sabdab: the structural antibody database}.
\bjtitle{Nucleic acids research}
\bvolume{42}(\bissue{D1}),
\bfpage{1140}--\blpage{1146}
(\byear{2014})
\end{barticle}
\endbibitem

\bibitem[\protect\citeauthoryear{}{}]{noauthor_tcr-epitope_nodate}
\begin{botherref}
{TCR}-Epitope Binding.
\url{https://tdcommons.ai/multi_pred_tasks/tcrepitope}
Accessed 2024-10-02
\end{botherref}
\endbibitem

\bibitem[\protect\citeauthoryear{Jankauskait{\.e}
  et~al.}{2019}]{jankauskaite2019skempi}
\begin{barticle}
\bauthor{\bsnm{Jankauskait{\.e}}, \binits{J.}},
\bauthor{\bsnm{Jim{\'e}nez-Garc{\'\i}a}, \binits{B.}},
\bauthor{\bsnm{Dapk{\=u}nas}, \binits{J.}},
\bauthor{\bsnm{Fern{\'a}ndez-Recio}, \binits{J.}},
\bauthor{\bsnm{Moal}, \binits{I.H.}}:
\batitle{Skempi 2.0: an updated benchmark of changes in protein--protein
  binding energy, kinetics and thermodynamics upon mutation}.
\bjtitle{Bioinformatics}
\bvolume{35}(\bissue{3}),
\bfpage{462}--\blpage{469}
(\byear{2019})
\end{barticle}
\endbibitem

\bibitem[\protect\citeauthoryear{Liu et~al.}{2023}]{liu2023persistent}
\begin{barticle}
\bauthor{\bsnm{Liu}, \binits{X.}},
\bauthor{\bsnm{Feng}, \binits{H.}},
\bauthor{\bsnm{L{\"u}}, \binits{Z.}},
\bauthor{\bsnm{Xia}, \binits{K.}}:
\batitle{Persistent tor-algebra for protein--protein interaction analysis}.
\bjtitle{Briefings in Bioinformatics}
\bvolume{24}(\bissue{2}),
\bfpage{046}
(\byear{2023})
\end{barticle}
\endbibitem

\bibitem[\protect\citeauthoryear{Wang et~al.}{2020}]{wang2020topology}
\begin{barticle}
\bauthor{\bsnm{Wang}, \binits{M.}},
\bauthor{\bsnm{Cang}, \binits{Z.}},
\bauthor{\bsnm{Wei}, \binits{G.-W.}}:
\batitle{A topology-based network tree for the prediction of protein--protein
  binding affinity changes following mutation}.
\bjtitle{Nature Machine Intelligence}
\bvolume{2}(\bissue{2}),
\bfpage{116}--\blpage{123}
(\byear{2020})
\end{barticle}
\endbibitem

\bibitem[\protect\citeauthoryear{Guo and Yamaguchi}{2022}]{guo2022machine}
\begin{barticle}
\bauthor{\bsnm{Guo}, \binits{Z.}},
\bauthor{\bsnm{Yamaguchi}, \binits{R.}}:
\batitle{Machine learning methods for protein-protein binding affinity
  prediction in protein design}.
\bjtitle{Frontiers in Bioinformatics}
\bvolume{2},
\bfpage{1065703}
(\byear{2022})
\end{barticle}
\endbibitem

\bibitem[\protect\citeauthoryear{Berman et~al.}{2000}]{10.1093/nar/28.1.235}
\begin{barticle}
\bauthor{\bsnm{Berman}, \binits{H.M.}},
\bauthor{\bsnm{Westbrook}, \binits{J.}},
\bauthor{\bsnm{Feng}, \binits{Z.}},
\bauthor{\bsnm{Gilliland}, \binits{G.}},
\bauthor{\bsnm{Bhat}, \binits{T.N.}},
\bauthor{\bsnm{Weissig}, \binits{H.}},
\bauthor{\bsnm{Shindyalov}, \binits{I.N.}},
\bauthor{\bsnm{Bourne}, \binits{P.E.}}:
\batitle{{The Protein Data Bank}}.
\bjtitle{Nucleic Acids Research}
\bvolume{28}(\bissue{1}),
\bfpage{235}--\blpage{242}
(\byear{2000})
\doiurl{10.1093/nar/28.1.235}
{\href{https://arxiv.org/abs/https://academic.oup.com/nar/article-pdf/28/1/235/9895144/280235.pdf}{{https://academic.oup.com/nar/article-pdf/28/1/235/9895144/280235.pdf}}}
\end{barticle}
\endbibitem

\bibitem[\protect\citeauthoryear{Gilson et~al.}{2015}]{bindingdb}
\begin{barticle}
\bauthor{\bsnm{Gilson}, \binits{M.K.}},
\bauthor{\bsnm{Liu}, \binits{T.}},
\bauthor{\bsnm{Baitaluk}, \binits{M.}},
\bauthor{\bsnm{Nicola}, \binits{G.}},
\bauthor{\bsnm{Hwang}, \binits{L.}},
\bauthor{\bsnm{Chong}, \binits{J.}}:
\batitle{{BindingDB} in 2015: A public database for medicinal chemistry,
  computational chemistry and systems pharmacology}.
\bjtitle{Nucleic Acids Res}
\bvolume{44}(\bissue{D1}),
\bfpage{1045}--\blpage{53}
(\byear{2015})
\end{barticle}
\endbibitem

\bibitem[\protect\citeauthoryear{Mason et~al.}{2021}]{mason_optimization_2021}
\begin{barticle}
\bauthor{\bsnm{Mason}, \binits{D.M.}},
\bauthor{\bsnm{Friedensohn}, \binits{S.}},
\bauthor{\bsnm{Weber}, \binits{C.R.}},
\bauthor{\bsnm{Jordi}, \binits{C.}},
\bauthor{\bsnm{Wagner}, \binits{B.}},
\bauthor{\bsnm{Meng}, \binits{S.M.}},
\bauthor{\bsnm{Ehling}, \binits{R.A.}},
\bauthor{\bsnm{Bonati}, \binits{L.}},
\bauthor{\bsnm{Dahinden}, \binits{J.}},
\bauthor{\bsnm{Gainza}, \binits{P.}},
\bauthor{\bsnm{Correia}, \binits{B.E.}},
\bauthor{\bsnm{Reddy}, \binits{S.T.}}:
\batitle{Optimization of therapeutic antibodies by predicting antigen
  specificity from antibody sequence via deep learning}.
\bjtitle{Nature Biomedical Engineering}
\bvolume{5}(\bissue{6}),
\bfpage{600}--\blpage{612}
(\byear{2021})
\doiurl{10.1038/s41551-021-00699-9} .
\bcomment{169 citations (Semantic Scholar/DOI) [2024-10-02] Number: 6
  Publisher: Nature Publishing Group}.
Accessed 2024-10-02
\end{barticle}
\endbibitem

\bibitem[\protect\citeauthoryear{Bennett
  et~al.}{2025}]{Bennett2024.03.14.585103}
\begin{barticle}
\bauthor{\bsnm{Bennett}, \binits{N.R.}},
\bauthor{\bsnm{Watson}, \binits{J.L.}},
\bauthor{\bsnm{Ragotte}, \binits{R.J.}},
\bauthor{\bsnm{Borst}, \binits{A.J.}},
\bauthor{\bsnm{See}, \binits{D.L.}},
\bauthor{\bsnm{Weidle}, \binits{C.}},
\bauthor{\bsnm{Biswas}, \binits{R.}},
\bauthor{\bsnm{Yu}, \binits{Y.}},
\bauthor{\bsnm{Shrock}, \binits{E.L.}},
\bauthor{\bsnm{Ault}, \binits{R.}},
\bauthor{\bsnm{Leung}, \binits{P.J.Y.}},
\bauthor{\bsnm{Huang}, \binits{B.}},
\bauthor{\bsnm{Goreshnik}, \binits{I.}},
\bauthor{\bsnm{Tam}, \binits{J.}},
\bauthor{\bsnm{Carr}, \binits{K.D.}},
\bauthor{\bsnm{Singer}, \binits{B.}},
\bauthor{\bsnm{Criswell}, \binits{C.}},
\bauthor{\bsnm{Wicky}, \binits{B.I.M.}},
\bauthor{\bsnm{Vafeados}, \binits{D.}},
\bauthor{\bsnm{Sanchez}, \binits{M.G.}},
\bauthor{\bsnm{Kim}, \binits{H.M.}},
\bauthor{\bsnm{V{\'a}zquez~Torres}, \binits{S.}},
\bauthor{\bsnm{Chan}, \binits{S.}},
\bauthor{\bsnm{Sun}, \binits{S.M.}},
\bauthor{\bsnm{Spear}, \binits{T.}},
\bauthor{\bsnm{Sun}, \binits{Y.}},
\bauthor{\bsnm{O{\textquoteright}Reilly}, \binits{K.}},
\bauthor{\bsnm{Maris}, \binits{J.M.}},
\bauthor{\bsnm{Sgourakis}, \binits{N.G.}},
\bauthor{\bsnm{Melnyk}, \binits{R.A.}},
\bauthor{\bsnm{Liu}, \binits{C.C.}},
\bauthor{\bsnm{Baker}, \binits{D.}}:
\batitle{Atomically accurate de novo design of antibodies with rfdiffusion}.
\bjtitle{bioRxiv}
(\byear{2025})
\doiurl{10.1101/2024.03.14.585103}
{\href{https://arxiv.org/abs/https://www.biorxiv.org/content/early/2025/02/28/2024.03.14.585103.full.pdf}{{https://www.biorxiv.org/content/early/2025/02/28/2024.03.14.585103.full.pdf}}}
\end{barticle}
\endbibitem

\bibitem[\protect\citeauthoryear{Yin and Pierce}{2024}]{af_ab_modeling}
\begin{barticle}
\bauthor{\bsnm{Yin}, \binits{R.}},
\bauthor{\bsnm{Pierce}, \binits{B.G.}}:
\batitle{Evaluation of alphafold antibody–antigen modeling with implications
  for improving predictive accuracy}.
\bjtitle{Protein Science}
\bvolume{33}(\bissue{1}),
\bfpage{4865}
(\byear{2024})
\doiurl{10.1002/pro.4865}
{\href{https://arxiv.org/abs/https://onlinelibrary.wiley.com/doi/pdf/10.1002/pro.4865}{{https://onlinelibrary.wiley.com/doi/pdf/10.1002/pro.4865}}}
\end{barticle}
\endbibitem

\bibitem[\protect\citeauthoryear{Schneider et~al.}{2021}]{sabdab-nano}
\begin{barticle}
\bauthor{\bsnm{Schneider}, \binits{C.}},
\bauthor{\bsnm{Raybould}, \binits{M.I.J.}},
\bauthor{\bsnm{Deane}, \binits{C.M.}}:
\batitle{Sabdab in the age of biotherapeutics: updates including sabdab-nano,
  the nanobody structure tracker}.
\bjtitle{Nucleic Acids Research}
\bvolume{50}(\bissue{D1}),
\bfpage{1368}--\blpage{1372}
(\byear{2021})
\doiurl{10.1093/nar/gkab1050}
{\href{https://arxiv.org/abs/https://academic.oup.com/nar/article-pdf/50/D1/D1368/42058451/gkab1050.pdf}{{https://academic.oup.com/nar/article-pdf/50/D1/D1368/42058451/gkab1050.pdf}}}
\end{barticle}
\endbibitem

\bibitem[\protect\citeauthoryear{Consortium}{2022}]{uniprot}
\begin{barticle}
\bauthor{\bsnm{Consortium}, \binits{T.U.}}:
\batitle{{UniProt: the Universal Protein Knowledgebase in 2023}}.
\bjtitle{Nucleic Acids Research}
\bvolume{51}(\bissue{D1}),
\bfpage{523}--\blpage{531}
(\byear{2022})
\doiurl{10.1093/nar/gkac1052}
{\href{https://arxiv.org/abs/https://academic.oup.com/nar/article-pdf/51/D1/D523/48441158/gkac1052.pdf}{{https://academic.oup.com/nar/article-pdf/51/D1/D523/48441158/gkac1052.pdf}}}
\end{barticle}
\endbibitem

\bibitem[\protect\citeauthoryear{Madani et~al.}{2020}]{madani2020progen}
\begin{botherref}
\oauthor{\bsnm{Madani}, \binits{A.}},
\oauthor{\bsnm{McCann}, \binits{B.}},
\oauthor{\bsnm{Naik}, \binits{N.}},
\oauthor{\bsnm{Keskar}, \binits{N.S.}},
\oauthor{\bsnm{Anand}, \binits{N.}},
\oauthor{\bsnm{Eguchi}, \binits{R.R.}},
\oauthor{\bsnm{Huang}, \binits{P.-S.}},
\oauthor{\bsnm{Socher}, \binits{R.}}:
Progen: Language modeling for protein generation.
arXiv preprint arXiv:2004.03497
(2020)
\end{botherref}
\endbibitem

\bibitem[\protect\citeauthoryear{Bunne et~al.}{2024}]{bunne2024build}
\begin{barticle}
\bauthor{\bsnm{Bunne}, \binits{C.}},
\bauthor{\bsnm{Roohani}, \binits{Y.}},
\bauthor{\bsnm{Rosen}, \binits{Y.}},
\bauthor{\bsnm{Gupta}, \binits{A.}},
\bauthor{\bsnm{Zhang}, \binits{X.}},
\bauthor{\bsnm{Roed}, \binits{M.}},
\bauthor{\bsnm{Alexandrov}, \binits{T.}},
\bauthor{\bsnm{AlQuraishi}, \binits{M.}},
\bauthor{\bsnm{Brennan}, \binits{P.}},
\bauthor{\bsnm{Burkhardt}, \binits{D.B.}}, \betal:
\batitle{How to build the virtual cell with artificial intelligence: Priorities
  and opportunities}.
\bjtitle{Cell}
\bvolume{187}(\bissue{25}),
\bfpage{7045}--\blpage{7063}
(\byear{2024})
\end{barticle}
\endbibitem

\bibitem[\protect\citeauthoryear{Song et~al.}{2024}]{song2024toward}
\begin{botherref}
\oauthor{\bsnm{Song}, \binits{L.}},
\oauthor{\bsnm{Segal}, \binits{E.}},
\oauthor{\bsnm{Xing}, \binits{E.}}:
Toward ai-driven digital organism: Multiscale foundation models for predicting,
  simulating and programming biology at all levels.
arXiv preprint arXiv:2412.06993
(2024)
\end{botherref}
\endbibitem

\bibitem[\protect\citeauthoryear{Gridach et~al.}{2025}]{gridach2025agentic}
\begin{botherref}
\oauthor{\bsnm{Gridach}, \binits{M.}},
\oauthor{\bsnm{Nanavati}, \binits{J.}},
\oauthor{\bsnm{Abidine}, \binits{K.Z.E.}},
\oauthor{\bsnm{Mendes}, \binits{L.}},
\oauthor{\bsnm{Mack}, \binits{C.}}:
Agentic ai for scientific discovery: A survey of progress, challenges, and
  future directions.
arXiv preprint arXiv:2503.08979
(2025)
\end{botherref}
\endbibitem

\bibitem[\protect\citeauthoryear{Ramos et~al.}{2025}]{ramos2025review}
\begin{botherref}
\oauthor{\bsnm{Ramos}, \binits{M.C.}},
\oauthor{\bsnm{Collison}, \binits{C.J.}},
\oauthor{\bsnm{White}, \binits{A.D.}}:
A review of large language models and autonomous agents in chemistry.
Chemical Science
(2025)
\end{botherref}
\endbibitem

\bibitem[\protect\citeauthoryear{Vaswani et~al.}{2017}]{vaswani2017attention}
\begin{botherref}
\oauthor{\bsnm{Vaswani}, \binits{A.}},
\oauthor{\bsnm{Shazeer}, \binits{N.}},
\oauthor{\bsnm{Parmar}, \binits{N.}},
\oauthor{\bsnm{Uszkoreit}, \binits{J.}},
\oauthor{\bsnm{Jones}, \binits{L.}},
\oauthor{\bsnm{Gomez}, \binits{A.N.}},
\oauthor{\bsnm{Kaiser}, \binits{L.}},
\oauthor{\bsnm{Polosukhin}, \binits{I.}}:
Attention is all you need.
Advances in Neural Information Processing Systems
(2017)
\end{botherref}
\endbibitem

\bibitem[\protect\citeauthoryear{Raffel et~al.}{2020}]{t5}
\begin{barticle}
\bauthor{\bsnm{Raffel}, \binits{C.}},
\bauthor{\bsnm{Shazeer}, \binits{N.}},
\bauthor{\bsnm{Roberts}, \binits{A.}},
\bauthor{\bsnm{Lee}, \binits{K.}},
\bauthor{\bsnm{Narang}, \binits{S.}},
\bauthor{\bsnm{Matena}, \binits{M.}},
\bauthor{\bsnm{Zhou}, \binits{Y.}},
\bauthor{\bsnm{Li}, \binits{W.}},
\bauthor{\bsnm{Liu}, \binits{P.J.}}:
\batitle{Exploring the limits of transfer learning with a unified text-to-text
  transformer}.
\bjtitle{Journal of Machine Learning Research}
\bvolume{21}(\bissue{140}),
\bfpage{1}--\blpage{67}
(\byear{2020})
\end{barticle}
\endbibitem

\bibitem[\protect\citeauthoryear{Chen et~al.}{2024}]{XTrimoPGLM}
\begin{botherref}
\oauthor{\bsnm{Chen}, \binits{B.}},
\oauthor{\bsnm{Cheng}, \binits{X.}},
\oauthor{\bsnm{Li}, \binits{P.}},
\oauthor{\bsnm{Geng}, \binits{Y.-a.}},
\oauthor{\bsnm{Gong}, \binits{J.}},
\oauthor{\bsnm{Li}, \binits{S.}},
\oauthor{\bsnm{Bei}, \binits{Z.}},
\oauthor{\bsnm{Tan}, \binits{X.}},
\oauthor{\bsnm{Wang}, \binits{B.}},
\oauthor{\bsnm{Zeng}, \binits{X.}},
\oauthor{\bsnm{Liu}, \binits{C.}},
\oauthor{\bsnm{Zeng}, \binits{A.}},
\oauthor{\bsnm{Dong}, \binits{Y.}},
\oauthor{\bsnm{Tang}, \binits{J.}},
\oauthor{\bsnm{Song}, \binits{L.}}:
xTrimoPGLM: Unified 100B-Scale Pre-trained Transformer for Deciphering the
  Language of Protein
(2024).
\url{https://arxiv.org/abs/2401.06199}
\end{botherref}
\endbibitem

\bibitem[\protect\citeauthoryear{Pei et~al.}{2023}]{pei2023biot5}
\begin{botherref}
\oauthor{\bsnm{Pei}, \binits{Q.}},
\oauthor{\bsnm{Zhang}, \binits{W.}},
\oauthor{\bsnm{Zhu}, \binits{J.}},
\oauthor{\bsnm{Wu}, \binits{K.}},
\oauthor{\bsnm{Gao}, \binits{K.}},
\oauthor{\bsnm{Wu}, \binits{L.}},
\oauthor{\bsnm{Xia}, \binits{Y.}},
\oauthor{\bsnm{Yan}, \binits{R.}}:
Biot5: Enriching cross-modal integration in biology with chemical knowledge and
  natural language associations.
arXiv preprint arXiv:2310.07276
(2023)
\end{botherref}
\endbibitem

\bibitem[\protect\citeauthoryear{Pei et~al.}{2024}]{pei2024biot5+}
\begin{botherref}
\oauthor{\bsnm{Pei}, \binits{Q.}},
\oauthor{\bsnm{Wu}, \binits{L.}},
\oauthor{\bsnm{Gao}, \binits{K.}},
\oauthor{\bsnm{Liang}, \binits{X.}},
\oauthor{\bsnm{Fang}, \binits{Y.}},
\oauthor{\bsnm{Zhu}, \binits{J.}},
\oauthor{\bsnm{Xie}, \binits{S.}},
\oauthor{\bsnm{Qin}, \binits{T.}},
\oauthor{\bsnm{Yan}, \binits{R.}}:
Biot5+: Towards generalized biological understanding with iupac integration and
  multi-task tuning.
arXiv preprint arXiv:2402.17810
(2024)
\end{botherref}
\endbibitem

\bibitem[\protect\citeauthoryear{Touvron et~al.}{2023}]{llama}
\begin{botherref}
\oauthor{\bsnm{Touvron}, \binits{H.}},
\oauthor{\bsnm{Lavril}, \binits{T.}},
\oauthor{\bsnm{Izacard}, \binits{G.}},
\oauthor{\bsnm{Martinet}, \binits{X.}},
\oauthor{\bsnm{Lachaux}, \binits{M.-A.}},
\oauthor{\bsnm{Lacroix}, \binits{T.}},
\oauthor{\bsnm{Rozière}, \binits{B.}},
\oauthor{\bsnm{Goyal}, \binits{N.}},
\oauthor{\bsnm{Hambro}, \binits{E.}},
\oauthor{\bsnm{Azhar}, \binits{F.}},
\oauthor{\bsnm{Rodriguez}, \binits{A.}},
\oauthor{\bsnm{Joulin}, \binits{A.}},
\oauthor{\bsnm{Grave}, \binits{E.}},
\oauthor{\bsnm{Lample}, \binits{G.}}:
LLaMA: Open and Efficient Foundation Language Models
(2023).
\url{https://arxiv.org/abs/2302.13971}
\end{botherref}
\endbibitem

\bibitem[\protect\citeauthoryear{Granite~Team}{2024}]{granite}
\begin{botherref}
\oauthor{\bsnm{Granite~Team}, \binits{I.}}:
Granite 3.0 Language Models
(2024).
\url{https://github.com/ibm-granite/granite-3.0-language-models/}
\end{botherref}
\endbibitem

\bibitem[\protect\citeauthoryear{Born and
  Manica}{2023}]{regression_transformers}
\begin{barticle}
\bauthor{\bsnm{Born}, \binits{J.}},
\bauthor{\bsnm{Manica}, \binits{M.}}:
\batitle{Regression transformer enables concurrent sequence regression and
  generation for molecular language modelling}.
\bjtitle{Nature Machine Intelligence}
\bvolume{5}(\bissue{4}),
\bfpage{432}--\blpage{444}
(\byear{2023})
\doiurl{10.1038/s42256-023-00639-z}
\end{barticle}
\endbibitem

\bibitem[\protect\citeauthoryear{Zausinger et~al.}{2024}]{regress_dont_guess}
\begin{botherref}
\oauthor{\bsnm{Zausinger}, \binits{J.}},
\oauthor{\bsnm{Pennig}, \binits{L.}},
\oauthor{\bsnm{Chlodny}, \binits{K.}},
\oauthor{\bsnm{Limbach}, \binits{V.}},
\oauthor{\bsnm{Ketteler}, \binits{A.}},
\oauthor{\bsnm{Prein}, \binits{T.}},
\oauthor{\bsnm{Singh}, \binits{V.M.}},
\oauthor{\bsnm{Danziger}, \binits{M.M.}},
\oauthor{\bsnm{Born}, \binits{J.}}:
Regress, Don't Guess -- A Regression-like Loss on Number Tokens for Language
  Models
(2024).
\url{https://arxiv.org/abs/2411.02083}
\end{botherref}
\endbibitem

\bibitem[\protect\citeauthoryear{Hayes et~al.}{2024}]{esm3}
\begin{barticle}
\bauthor{\bsnm{Hayes}, \binits{T.}},
\bauthor{\bsnm{Rao}, \binits{R.}},
\bauthor{\bsnm{Akin}, \binits{H.}},
\bauthor{\bsnm{Sofroniew}, \binits{N.J.}},
\bauthor{\bsnm{Oktay}, \binits{D.}},
\bauthor{\bsnm{Lin}, \binits{Z.}},
\bauthor{\bsnm{Verkuil}, \binits{R.}},
\bauthor{\bsnm{Tran}, \binits{V.Q.}},
\bauthor{\bsnm{Deaton}, \binits{J.}},
\bauthor{\bsnm{Wiggert}, \binits{M.}},
\bauthor{\bsnm{Badkundri}, \binits{R.}},
\bauthor{\bsnm{Shafkat}, \binits{I.}},
\bauthor{\bsnm{Gong}, \binits{J.}},
\bauthor{\bsnm{Derry}, \binits{A.}},
\bauthor{\bsnm{Molina}, \binits{R.S.}},
\bauthor{\bsnm{Thomas}, \binits{N.}},
\bauthor{\bsnm{Khan}, \binits{Y.}},
\bauthor{\bsnm{Mishra}, \binits{C.}},
\bauthor{\bsnm{Kim}, \binits{C.}},
\bauthor{\bsnm{Bartie}, \binits{L.J.}},
\bauthor{\bsnm{Nemeth}, \binits{M.}},
\bauthor{\bsnm{Hsu}, \binits{P.D.}},
\bauthor{\bsnm{Sercu}, \binits{T.}},
\bauthor{\bsnm{Candido}, \binits{S.}},
\bauthor{\bsnm{Rives}, \binits{A.}}:
\batitle{Simulating 500 million years of evolution with a language model}.
\bjtitle{bioRxiv}
(\byear{2024})
\doiurl{10.1101/2024.07.01.600583}
{\href{https://arxiv.org/abs/https://www.biorxiv.org/content/early/2024/07/02/2024.07.01.600583.full.pdf}{{https://www.biorxiv.org/content/early/2024/07/02/2024.07.01.600583.full.pdf}}}
\end{barticle}
\endbibitem

\bibitem[\protect\citeauthoryear{van~den Oord et~al.}{2018}]{vqvae}
\begin{botherref}
\oauthor{\bsnm{Oord}, \binits{A.}},
\oauthor{\bsnm{Vinyals}, \binits{O.}},
\oauthor{\bsnm{Kavukcuoglu}, \binits{K.}}:
Neural Discrete Representation Learning
(2018).
\url{https://arxiv.org/abs/1711.00937}
\end{botherref}
\endbibitem

\bibitem[\protect\citeauthoryear{Olsen et~al.}{2022}]{oas}
\begin{barticle}
\bauthor{\bsnm{Olsen}, \binits{T.H.}},
\bauthor{\bsnm{Boyles}, \binits{F.}},
\bauthor{\bsnm{Deane}, \binits{C.M.}}:
\batitle{Observed antibody space: A diverse database of cleaned, annotated, and
  translated unpaired and paired antibody sequences}.
\bjtitle{Protein Science}
\bvolume{31}(\bissue{1}),
\bfpage{141}--\blpage{146}
(\byear{2022})
\doiurl{10.1002/pro.4205}
{\href{https://arxiv.org/abs/https://onlinelibrary.wiley.com/doi/pdf/10.1002/pro.4205}{{https://onlinelibrary.wiley.com/doi/pdf/10.1002/pro.4205}}}
\end{barticle}
\endbibitem

\bibitem[\protect\citeauthoryear{Tingle et~al.}{2023}]{zinc22}
\begin{barticle}
\bauthor{\bsnm{Tingle}, \binits{B.I.}},
\bauthor{\bsnm{Tang}, \binits{K.G.}},
\bauthor{\bsnm{Castanon}, \binits{M.}},
\bauthor{\bsnm{Gutierrez}, \binits{J.J.}},
\bauthor{\bsnm{Khurelbaatar}, \binits{M.}},
\bauthor{\bsnm{Dandarchuluun}, \binits{C.}},
\bauthor{\bsnm{Moroz}, \binits{Y.S.}},
\bauthor{\bsnm{Irwin}, \binits{J.J.}}:
\batitle{Zinc-22-a free multi-billion-scale database of tangible compounds for
  ligand discovery}.
\bjtitle{Journal of Chemical Information and Modeling}
\bvolume{63}(\bissue{4}),
\bfpage{1166}--\blpage{1176}
(\byear{2023})
\doiurl{10.1021/acs.jcim.2c01253}
{\href{https://arxiv.org/abs/https://doi.org/10.1021/acs.jcim.2c01253}{{https://doi.org/10.1021/acs.jcim.2c01253}}}.
\bcomment{PMID: 36790087}
\end{barticle}
\endbibitem

\bibitem[\protect\citeauthoryear{Kim et~al.}{2022}]{pubchem}
\begin{barticle}
\bauthor{\bsnm{Kim}, \binits{S.}},
\bauthor{\bsnm{Chen}, \binits{J.}},
\bauthor{\bsnm{Cheng}, \binits{T.}},
\bauthor{\bsnm{Gindulyte}, \binits{A.}},
\bauthor{\bsnm{He}, \binits{J.}},
\bauthor{\bsnm{He}, \binits{S.}},
\bauthor{\bsnm{Li}, \binits{Q.}},
\bauthor{\bsnm{Shoemaker}, \binits{B.A.}},
\bauthor{\bsnm{Thiessen}, \binits{P.A.}},
\bauthor{\bsnm{Yu}, \binits{B.}},
\bauthor{\bsnm{Zaslavsky}, \binits{L.}},
\bauthor{\bsnm{Zhang}, \binits{J.}},
\bauthor{\bsnm{Bolton}, \binits{E.E.}}:
\batitle{{PubChem 2023 update}}.
\bjtitle{Nucleic Acids Research}
\bvolume{51}(\bissue{D1}),
\bfpage{1373}--\blpage{1380}
(\byear{2022})
\doiurl{10.1093/nar/gkac956}
{\href{https://arxiv.org/abs/https://academic.oup.com/nar/article-pdf/51/D1/D1373/48441598/gkac956.pdf}{{https://academic.oup.com/nar/article-pdf/51/D1/D1373/48441598/gkac956.pdf}}}
\end{barticle}
\endbibitem

\bibitem[\protect\citeauthoryear{Biology et~al.}{2023}]{CELLxGENE}
\begin{botherref}
\oauthor{\bsnm{Biology}, \binits{C.S.-C.}},
\oauthor{\bsnm{Abdulla}, \binits{S.}},
\oauthor{\bsnm{Aevermann}, \binits{B.}},
\oauthor{\bsnm{Assis}, \binits{P.}},
\oauthor{\bsnm{Badajoz}, \binits{S.}},
\oauthor{\bsnm{Bell}, \binits{S.M.}},
\oauthor{\bsnm{Bezzi}, \binits{E.}},
\oauthor{\bsnm{Cakir}, \binits{B.}},
\oauthor{\bsnm{Chaffer}, \binits{J.}},
\oauthor{\bsnm{Chambers}, \binits{S.}}, et al.:
Cz cellxgene discover: A single-cell data platform for scalable exploration,
  analysis and modeling of aggregated data.
BioRxiv,
2023--10
(2023)
\end{botherref}
\endbibitem

\bibitem[\protect\citeauthoryear{Szklarczyk et~al.}{2022}]{string}
\begin{barticle}
\bauthor{\bsnm{Szklarczyk}, \binits{D.}},
\bauthor{\bsnm{Kirsch}, \binits{R.}},
\bauthor{\bsnm{Koutrouli}, \binits{M.}},
\bauthor{\bsnm{Nastou}, \binits{K.}},
\bauthor{\bsnm{Mehryary}, \binits{F.}},
\bauthor{\bsnm{Hachilif}, \binits{R.}},
\bauthor{\bsnm{Gable}, \binits{A.L.}},
\bauthor{\bsnm{Fang}, \binits{T.}},
\bauthor{\bsnm{Doncheva}, \binits{N.}},
\bauthor{\bsnm{Pyysalo}, \binits{S.}},
\bauthor{\bsnm{Bork}, \binits{P.}},
\bauthor{\bsnm{Jensen}, \binits{L.}},
\bauthor{\bsnm{von Mering}, \binits{C.}}:
\batitle{{The STRING database in 2023: protein–protein association networks
  and functional enrichment analyses for any sequenced genome of interest}}.
\bjtitle{Nucleic Acids Research}
\bvolume{51}(\bissue{D1}),
\bfpage{638}--\blpage{646}
(\byear{2022})
\doiurl{10.1093/nar/gkac1000}
{\href{https://arxiv.org/abs/https://academic.oup.com/nar/article-pdf/51/D1/D638/48440966/gkac1000.pdf}{{https://academic.oup.com/nar/article-pdf/51/D1/D638/48440966/gkac1000.pdf}}}
\end{barticle}
\endbibitem

\bibitem[\protect\citeauthoryear{Xiong et~al.}{2017}]{xiong2017bindprofx}
\begin{barticle}
\bauthor{\bsnm{Xiong}, \binits{P.}},
\bauthor{\bsnm{Zhang}, \binits{C.}},
\bauthor{\bsnm{Zheng}, \binits{W.}},
\bauthor{\bsnm{Zhang}, \binits{Y.}}:
\batitle{Bindprofx: assessing mutation-induced binding affinity change by
  protein interface profiles with pseudo-counts}.
\bjtitle{Journal of molecular biology}
\bvolume{429}(\bissue{3}),
\bfpage{426}--\blpage{434}
(\byear{2017})
\end{barticle}
\endbibitem

\bibitem[\protect\citeauthoryear{Jan}{}]{HER2_repo_2024}
\begin{botherref}
\oauthor{\bsnm{Jan}}:
Dahjan/{DMS}\_opt.
original-date: 2020-06-24T09:29:12Z.
\url{https://github.com/dahjan/DMS_opt}
Accessed 2024-10-27
\end{botherref}
\endbibitem

\bibitem[\protect\citeauthoryear{Suzek et~al.}{2015}]{uniref}
\begin{barticle}
\bauthor{\bsnm{Suzek}, \binits{B.E.}},
\bauthor{\bsnm{Wang}, \binits{Y.}},
\bauthor{\bsnm{Huang}, \binits{H.}},
\bauthor{\bsnm{McGarvey}, \binits{P.B.}},
\bauthor{\bsnm{Wu}, \binits{C.H.}},
\bauthor{\bsnm{{UniProt Consortium}}}:
\batitle{{UniRef} clusters: a comprehensive and scalable alternative for
  improving sequence similarity searches}.
\bjtitle{Bioinformatics}
\bvolume{31}(\bissue{6}),
\bfpage{926}--\blpage{932}
(\byear{2015})
\end{barticle}
\endbibitem

\bibitem[\protect\citeauthoryear{Golts et~al.}{2023}]{fuse}
\begin{barticle}
\bauthor{\bsnm{Golts}, \binits{A.}},
\bauthor{\bsnm{Raboh}, \binits{M.}},
\bauthor{\bsnm{Shoshan}, \binits{Y.}},
\bauthor{\bsnm{Polaczek}, \binits{S.}},
\bauthor{\bsnm{Rabinovici-Cohen}, \binits{S.}},
\bauthor{\bsnm{Hexter}, \binits{E.}}:
\batitle{Fusemedml: a framework for accelerated discovery in machine learning
  based biomedicine}.
\bjtitle{Journal of Open Source Software}
\bvolume{8}(\bissue{81}),
\bfpage{4943}
(\byear{2023})
\doiurl{10.21105/joss.04943}
\end{barticle}
\endbibitem

\end{thebibliography}


\newpage

\restylefloat{table}

\begin{appendices}

\setcounter{table}{0}  
\renewcommand\theHtable{Appendix.\thetable}

\renewcommand{\thetable}{S\arabic{table}}  

\newpage
\section{Architecture - Additional Details}\label{architecture_more_details}

\begin{figure}[h]
\centering

\includegraphics[width = 350pt]{arch_detailed_v8_svg-raw.pdf}

\caption{A prompt, consisting of both token IDs and scalars is processed and enters the encoder. Both the encoder and the decoder output logits which are used for classification loss. Additionally, the output of the encoder is sent to a (learned) scalars prediction head which allows the prediction scalars for any subset of the tokens, and is used in the regression loss. In this illustration, a single scalar input ("12.7") is being used, and a single scalar's outputs are predicted by the model ("97.2"). However, the method fully supports an arbitrary number of input scalars and outputs.}\label{fig_arch_detailed}
\end{figure}

One of the key aspects in the MAMMAL method is the built-in support for scalar inputs and outputs.
Figure \ref{fig_arch_detailed} illustrates how this is achieved.

A user prompt, usually expressed as a single text line, is processed into two input sequences: a. Input token IDs, which are a sequence of integer values representing tokens in the vocabulary, and b. a sequence of inputs scalars (by convention, containing NaNs for positions for which no input scalar is provided).

The input tokens IDs are transformed using a learned token embedding, and the input scalars are transformed using a learned linear transformation that projects each single scalar element into the model dimension (e.g, 768).

Both representations are added (not concatenated) and fed into the encoder stack. Using this approach, both tasks that use encoder-only mode and encoder-decoder mode benefit from the ability to get as input an arbitrary number of scalars (at most as many as the number of tokens that are being fed in ).

For scalars outputs (gene expression, binding free energy, etc.), the encoder stack has an additional prediction head, which outputs a scalar value for every input element. How to deal with scalar outputs in locations where there is no scalar label is the user's choice, but the default is to ignore those.

The support of scalar outputs in the encoder-decoder mode is an improvement that we intend to add in future generations of the model/method.

\newpage
\section{Prompt Syntax}\label{query_syntax_detailed}

A typical prompt is built as a combination of entities of the following types:

\begin{itemize}
\item \sequenceWrp{Sequence}: A sequence of amino acids or other chemical representations, such as SMILES, which may encompass a full sequence or a specific region.
\item \molecule{Molecule}: A complete molecule, such as a protein chain or small molecule, which may contain multiple sequences corresponding to sub-regions within the molecule. Each molecule is initiated with two special tokens - a general token indicating the entity’s hierarchical level, \molecule{\token{MOLECULAR\_ENTITY}}, followed by a token specifying the molecule type (e.g., \molecule{\token{MOLECULAR\_ENTITY\_EPITOPE}}). Additionally, molecules can be marked with natural start and end tokens to denote instances where truncation has occurred, either in the original database or due to sequence length constraints.
\item \complex{MolecularSystem}: A quaternary structure consisting of multiple molecules, denoted by the \complex{\token{COMPLEX\_ENTITY}} special token.
\item \globalent{GlobalSystem}: A system comprising multiple interacting MolecularSystem entities.
\item Attribute: A representation of properties or interactions among the entities.
\end{itemize}
\noindent
\\
To support a wide range of tasks, MAMMAL utilizes special placeholder tokens of the form \token{SENTINEL\_ID\_?}. These tokens serve as positionally aware anchors in the decoder output and are designed to flexibly handle both structured and unstructured prediction targets. For complex generation tasks, such as sequence infilling, molecule editing, or antibody design, multiple sentinel tokens (e.g., \token{SENTINEL\_ID\_0}, \token{SENTINEL\_ID\_1} allow the model to generate multiple, disjoint segments of output. This enables fine-grained control over where the model should insert or complete information.

In simpler settings such as classification or regression, a single sentinel token (e.g., \token{SENTINEL\_ID\_0} is used to indicate the position where the model should output the prediction. This unified approach allows all tasks, whether discrete or continuous, generative or discriminative, to be framed consistently as a form of sequence-to-sequence learning, while preserving flexibility across domains. The use of sentinel tokens thus enables seamless task specification and modular decoding, making the model architecture adaptable to a broad spectrum of biomedical problems.\\

\begin{figure}[h]
\centering
\includegraphics[width = 350pt]{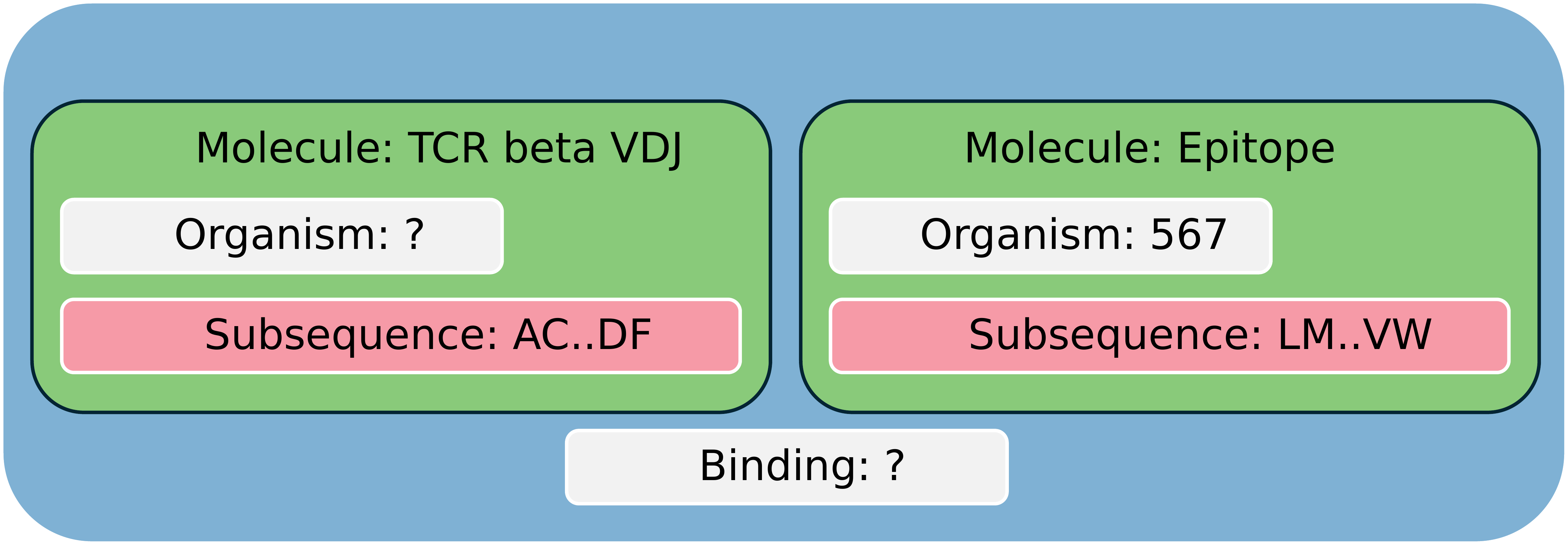}
\caption{Entity hierarchy for the task of binding prediction of two proteins, and organism prediction of the first one.}\label{fig_query_organism_1}
\end{figure}

\noindent
Example 1: Illustrated in Figure \ref{fig_query_organism_1}. Given two interacting molecules – variable region of a TCR beta chain of an unspecified organism and an epitope of organism 567, find whether they bind, and to which organism the first molecule belongs.:\\
\begin{compactitem} 
\item encoder inputs = 
\metatoken{\token{@TOKENIZER-TYPE=AA}} 
\globalent{\token{BINDING\_AFFINITY\_CLASS}\token{SENTINEL\_ID\_0}} \molecule{\token{MOLECULAR\_ENTITY}\token{MOLECULAR\_ENTITY\_TCR\_BETA\_VDJ}}\molecule{\token{ATTRIBUTE\_ORGANISM}\token{SENTINEL\_ID\_1}}
\sequenceWrp{\texttt{AC\dots DF}}\molecule{\token{MOLECULAR\_ENTITY}\token{MOLECULAR\_ENTITY\_EPITOPE}}\molecule{\token{ATTRIBUTE\_ORGANISM}\token{5}\token{6}\token{7}\token{8}}
\sequenceWrp{\texttt{LM\dots VW}}
\token{EOS}
\item labels = \token{@TOKENIZER-TYPE=AA}\token{SENTINEL\_ID\_0}\token{1}\token{SENTINEL\_ID\_1}\token{2}\token{3}\token{4}\token{EOS}
\end{compactitem} 
\bigskip\bigskip

\begin{figure}[h]
\centering
\includegraphics[width = 350pt]{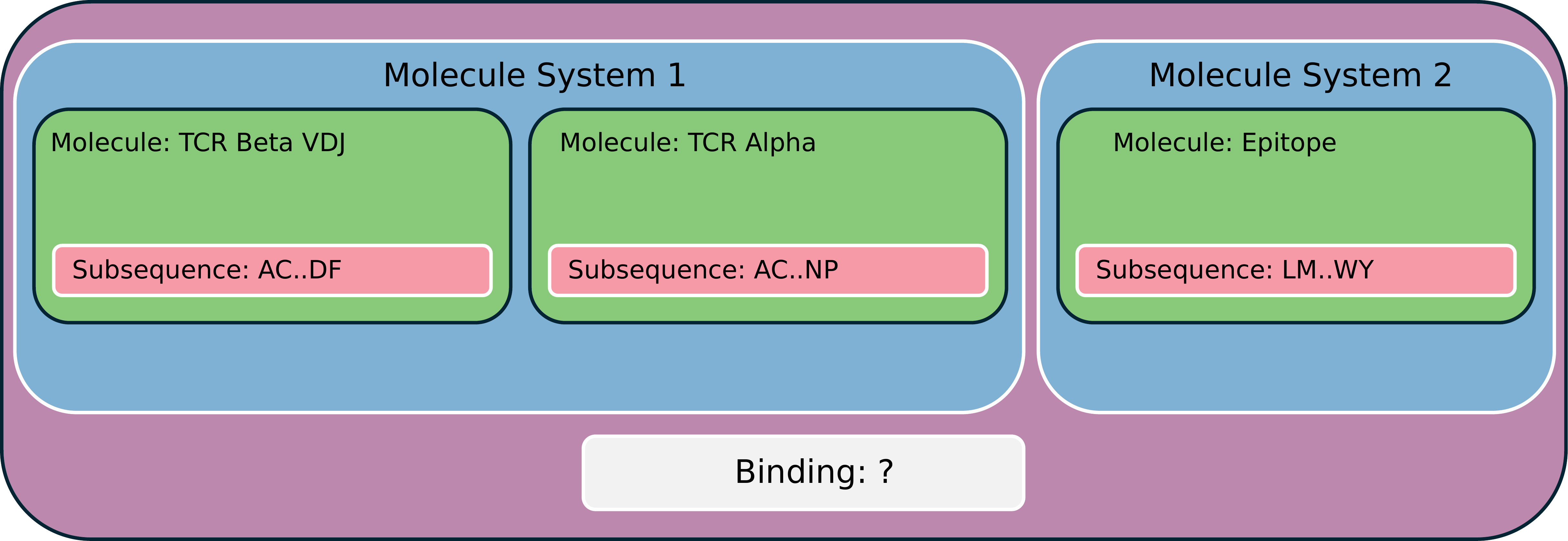}
\caption{Entity hierarchy for the task of binding prediction of a TCR and an epitope. "Molecule System 1" represents the TCR complex, "Molecule System 2" represents the antigen, and the entire prompt represents their interaction. }\label{fig_query_complex_2}
\end{figure}

\noindent
Example 2 : Illustrated in Figure \ref{fig_query_complex_2}. Given a complex entity, T-cell receptor consisting of alpha and beta chains, and an epitope, predict if they bind.:\\
\begin{compactitem} 

\item \raggedright encoder inputs = 
\metatoken{\token{@TOKENIZER-TYPE=AA}}\globalent{\token{BINDING\_AFFINITY\_CLASS}\token{SENTINEL\_ID\_0}}\\
\complex{\token{COMPLEX\_ENTITY}}\molecule{\token{MOLECULAR\_ENTITY}\token{MOLECULAR\_ENTITY\_TCR\_BETA\_VDJ}}\sequenceWrp{\texttt{AB\dots DF}}\molecule{\token{MOLECULAR\_ENTITY}\token{MOLECULAR\_ENTITY\_TCR\_ALPHA}}
\sequenceWrp{\texttt{AC\dots NP}}\\
\complex{\token{COMPLEX\_ENTITY}}
\molecule{\token{MOLECULAR\_ENTITY}\token{MOLECULAR\_ENTITY\_EPITOPE}}\sequenceWrp{\texttt{LM\dots WY}}
\token{EOS}
\item \raggedright labels = \token{@TOKENIZER-TYPE=AA}\token{SENTINEL\_ID\_0}\token{1}\token{EOS}\\
\end{compactitem}

\bigskip\bigskip
\noindent
Example: Given two binding chains – TCR beta chain and an epitope, unmask 3 spans within the beta chain:\\
\begin{compactitem} 
\item \raggedright labels = \token{@TOKENIZER-TYPE=AA}\token{SENTINEL\_ID\_0}AC\token{SENTINEL\_ID\_1}AD\token{SENTINEL\_ID\_2}CD\token{EOS}
\item \raggedright encoder inputs = \token{@TOKENIZER-TYPE=AA} \token{BINDING\_AFFINITY\_CLASS}\token{1}\token{MOLECULAR\_ENTITY}\token{MOLECULAR\_ENTITY\_EPITOPE}\token{SEQUENCE\_NATURAL\_START}LM\dots WY\token{SEQUENCE\_NATURAL\_END}\token{MOLECULAR\_ENTITY}\token{MOLECULAR\_ENTITY\_TCR\_BETA\_VDJ}\sequenceWrp{    \token{SENTINEL\_ID\_0}C\dots D\nolinebreak[4]\token{SENTINEL\_ID\_1}{DF}\token{SENTINEL\_ID\_2}{FDF}}\token{EOS}
\end{compactitem} 
\bigskip\bigskip
\noindent
Example 3: (for multitokenizer): DTI\\
\begin{compactitem} 
\item labels = \token{@TOKENIZER-TYPE=AA}\token{SENTINEL\_ID\_0}\token{1}\token{EOS}
\item \raggedright encoder inputs = \token{@TOKENIZER-TYPE=AA}\token{BINDING\_AFFINITY\_CLASS}\token{SENTINEL\_ID\_0}\token{@TOKENIZER-TYPE=SMILES@MAX-LEN=10}\token{MOLECULAR\_ENTITY}\token{MOLECULAR\_ENTITY\_SMALL\_MOLECULE}\sequenceWrp{CCC=CCC}
\token{@TOKENIZER-TYPE=AA@MAXLEN=15}\token{MOLECULAR\_ENTITY}\token{MOLECULAR\_ENTITY\_EPITOPE}\\ \sequenceWrp{LMNPQRSTUVWY}\token{EOS}
\end{compactitem} 
\bigskip\bigskip
\noindent

Examples showing how a prompt is created for several downstream tasks can be found in Table  \ref{table_task_queries}.

\subsection{Modular Tokenizer and Meta Tokens}

To support multiple modalities within a single prompt, we developed "Modular Tokenizer" which allows to utilization of different tokenizers within a single prompt by mapping tokens from different domains (like SMILES carbon "C" and amino acid cysteine "C") to the same ID space. 

We use meta tokens of the format \token{@TOKENIZER-TYPE=...} to indicate that everything following this meta token, up to the next meta token (or the end of the prompt) should be tokenized with the defined tokenizer. For example, \token{@TOKENIZER-TYPE=AA} tokenizes amino-acids, while \token{@TOKENIZER-TYPE=SMILES} can tokenize SMILES.
Since all of those "sub tokenizers" must exist in a single vocabulary space, our modular tokenizer orchestrates that, and provides mechanism to avoid conflicts and allow for sub-tokenizers to co-exist.
Additionally, we support additional instructions within a meta token, beyond just the expression of which (sub) tokenizer should be used.
For example, \token{@TOKENIZER-TYPE=AA@MAX-LEN=1000} allows users to restrict the maximum length of the tokenized sequence, which provides more granular control compared to only controlling the overall total max sequence tokenized length.
It is worth emphasizing that meta tokens, by themselves,  do not get tokenized into any token. They serve as instructions for the modular tokenizer.
Further details on the implementation can be found on \url{https://github.com/BiomedSciAI/fuse-med-ml/tree/master/fuse/data/tokenizers/modular_tokenizer}

\newpage
\section{Pretraining Details}\label{appendix_pretraining}
\subsection{Infrastructure}
\MAMMALMODEL model was trained on an OpenShift cluster. It was trained for three months over two nodes with 16 A100-80G GPUs. The training framework was implemented using FuseMedML \cite{fuse} and PyTorch, with distributed processing supported by PyTorch Fully Sharded Data Parallel (FSDP) for efficient parallelism.
 
\subsection{Hyperparameters}
We train \MAMMALMODEL using AdamW optimizer, with the following hyperparameters:  \(\beta1 = 0.9\), \(\beta2 = 0.999\). We use a weight decay of 0.01 and a gradient clipping norm of 1.0. We employ 2K warmup steps until reaching the maximum learning rate and utilize a cosine decay scheduler to decay LR to \(10\%\) of the maximum learning rate by the end of training.
The maximum sequence length was set per task to be effective yet efficient. When required, instead of naively truncating the end of a sequence, we first wrapped the sequence with special start and end tokens to provide a hint for the model as to whether the beginning or end of a sequence was truncated. Then we randomly cut a random sub-sequence with the required length.
Batch sizes were tuned per task given the maximum sequence length to maximize GPU memory utilization. 

\subsection{Datasets}
\MAMMALMODEL was pre-trained using six diverse datasets spanning multiple domains.\\
\textbf{UniRef90} \cite{uniref}, one of the clustered databases in UniProt \cite{uniprot} (UniProt Reference Clusters), groups protein sequences that share at least 90\% identity and 80\% sequence overlap with the longest sequence in each cluster (the seed sequence). This clustering approach reduces redundancy while preserving the diversity of functional sequences, providing a rich protein dataset.\\
\textbf{OAS} \cite{oas} (Observed Antibody Space) offers unpaired antibody sequences, specifically focusing on the variable regions of light or heavy chains. After filtering for sequences with complete variable domains and retaining only the sequences with standard amino acids, we finalized a dataset of approximately 650 million sequences. Each sequence is annotated with its chain type (heavy or light) and species information.\\
\textbf{STRING} \cite{string} is a database that integrates data from experimental findings, computational predictions, and text mining to describe protein-protein interactions. We curated a dataset of 390 million positive protein interaction pairs, considering only pairs that had a STRING confidence score above 500. Additionally, we curated 390 million pseudo-negative pairs by randomly matching proteins from the same species, resulting in a second dataset of 780 million samples.\\
\textbf{CELLxGENE} \cite{CELLxGENE}, a platform for single-cell transcriptomics data, was used to assemble a dataset of gene expression sequences from individual cells. After filtering for human samples labeled as “cell” in the 'suspension\_type' field, we curated a dataset of 30 million samples. 
The sequences for model training were created as ranked lists of gene names, similar to Geneformer~\cite{genformer}, but with several differences. 
Instead of normalizing by median expression and ranking the now unique values of expression count, we log-normalized and binned the samples, followed by alphabetic sorting to make the task well-defined.
The binning removes the ability of noisy reads to significantly affect the pretraining and alphabetic sorting ensures that for every masked token there is a unique correct answer.\\
For small-molecule data, we utilized two main sources: (1) \textbf{PubChem} \cite{pubchem}, a comprehensive chemical database maintained by NCBI, and (2) \textbf{ZINC22} \cite{zinc22}, a large library of drug-like molecules. From PubChem, we curated a subset of 80 million “drug-like” molecules, removing duplicates and following Lipinski's rule of five to ensure drug-likeness. Additionally, we sampled 120 million molecules from the ZINC22 database, focusing on small molecules with fewer than 30 heavy atoms to ensure dense coverage of “drug-like” chemical space. This led to a final dataset of 200 million small-molecule examples.

\subsection{Multitask Pretraining Approach}
\MAMMALMODEL was pre-trained on seven tasks simultaneously using a multitask learning approach. Gradients from each task were aggregated before updating the optimizer. This approach, combined with a custom query system, enables the model to learn from different tasks and domains during co-training.

\subsection{Pretraining Tasks}

\textbf{Language Model Tasks.}
Four language model tasks were defined: (1) amino-acid sequence representation of antibodies based on OAS database \cite{oas}; (2) amino-acid sequence representation of general proteins based on Uniref90 \cite{uniprot,uniref}; (3) SMILES representation of small molecules based on a mixture of PubChem \cite{pubchem} and  Zinc databases \cite{zinc22}; and (4) Geneformer format representations \cite{genformer} of cell genes based on CELLxGENE \cite{CELLxGENE}.  
In all language modeling tasks, we employed span-denoising (similar to T5 \cite{t5}) with a mean noise span length of 5 and a noise density of 0.15. Additionally, a special token was introduced per entity type to make the model aware of it (e.g., \token{MOLECULAR\_ENTITY\_TYPE\_ANTIBODY\_HEAVY\_CHAIN}). When sequences were available from different species, an additional special token was also introduced (e.g., \token{ATTRIBUTE\_ORGANISM\_HUMAN})

\textbf{Antibody Denoise.}
This task focuses on recovering corrupted antibodies, represented by amino acid sequences. The corrupted sequence is generated by first sampling a value \textit{t} from the range [1, 500], and then uniformly corrupting the amino acid tokens with a probability proportional to \textit{t}. The antibody sequences used in this task are sourced from the OAS (Observatory of Antibody Space) dataset.

\textbf{Protein-Protein Interaction.}
As part of the pretraining process, two tasks were defined for learning protein-protein interactions: a classification task and a generation task, both utilizing data from the STRING database \cite{string}. Interactions with a score higher than 500 are labeled as positive, while random pairs of proteins are treated as negative interactions. For the classification task, a balanced dataset comprising both positive and negative pairs is used. In the generation task, the model is trained on positive pairs only, where it learns to generate an interacting protein given an input protein.

\newpage
\section{Additional Results}\label{appendix_additional_results}

\suppressfloats[t]

\begingroup
\def\_{\allowbreak\textunderscore}
\def\={\allowbreak\texttt{=}}
\def\-{\allowbreak\texttt{-}}
\def\dotsbreak{\allowbreak\dots}

\newcommand{\tokenmeta}[1]{%
    \textcolor[HTML]{c66aba}{\textlangle\texttt{#1}\textrangle}\allowbreak
}
\renewcommand{\arraystretch}{1.5}

\begin{longtable}{|L{1.2cm} L{7cm} L{4.2cm}|}
\caption{Examples of Encoder Inputs and Decoder Outputs for Benchmarks} \label{table_task_queries} \\

\hline Bench. & Encoder input & Decoder label \\  \hline 
\endfirsthead

\multicolumn{3}{c}%
{{\bfseries \tablename\ \thetable{} -- continued from previous page}} \\

\hline Bench. & Encoder input & Decoder label \\  \hline 

\endhead

\hline \multicolumn{3}{|r|}{{Continued on next page}} \\ \hline
\endfoot

\hline \hline
\endlastfoot


%
Cell type & \metatoken{\token{@TOKENIZER-TYPE=GENE}}\token{MOLECULAR\_ENTITY}\token{MOLECULAR\_ENTITY\_CELL\_GENE\_EXPRESSION\_RANKED}\sequenceWrp{\texttt{[MALAT1][RPL10]\dots[ZNF136][ZNF514]}}\token{CELL\_TYPE\_CLASS}\highlight{\token{SENTINEL\_ID\_0}}\token{EOS} & \metatoken{\token{@TOKENIZER-TYPE=CELL\_ATTRIBUTES}}\highlight{\token{SENTINEL\_ID\_0}\sequenceWrp{[CL:0001062]}}\token{EOS} \\

BBBP & 
\metatoken{\token{@TOKENIZER-TYPE=SMILES}}\token{MOLECULAR\_ENTITY}\token{MOLECULAR\_ENTITY\_SMALL\_MOLECULE}\token{BBBP}\highlight{\token{SENTINEL\_ID\_0}}\metatoken{\token{@TOKENIZER-TYPE=SMILES@MAX-LEN=2100}}\sequenceWrp{\texttt{C(Cl)Cl}}\token{EOS} & 
\metatoken{\token{@TOKENIZER-TYPE=SMILES}}\highlight{\token{SENTINEL\_ID\_0}\token{1}}\token{EOS} \\

ClinTox Toxic & 
\metatoken{\token{@TOKENIZER-TYPE=SMILES}}\token{MOLECULAR\_ENTITY}\token{MOLECULAR\_ENTITY\_SMALL\_MOLECULE}\token{TOXICITY}\highlight{\token{SENTINEL\_ID\_0}}\metatoken{\token{@TOKENIZER-TYPE=SMILES@MAX-LEN=2100}}\sequenceWrp{\texttt{C\#CC1(CCCCC1)OC(=O)N}}\token{EOS} & 
\metatoken{\token{@TOKENIZER-TYPE=SMILES}}\highlight{\token{SENTINEL\_ID\_0}\token{0}}\token{EOS} \\

ClinTox FDA Approval & 
\metatoken{\token{@TOKENIZER-TYPE=SMILES}}\token{MOLECULAR\_ENTITY}\token{MOLECULAR\_ENTITY\_SMALL\_MOLECULE}\token{FDA\_APPR}\highlight{\token{SENTINEL\_ID\_0}}\metatoken{\token{@TOKENIZER-TYPE=SMILES@MAX-LEN=2100}}\sequenceWrp{\texttt{C\#CC1(CCCCC1)OC(=O)N}}\token{EOS} & 
\metatoken{\token{@TOKENIZER-TYPE=SMILES}}\highlight{\token{SENTINEL\_ID\_0}\token{1}}\token{EOS} \\

Cancer-Drug Response & \metatoken{\token{@TOKENIZER-TYPE=SCALARS\_LITERALS}}\highlight{\textless MASK \textgreater} \metatoken{\token{@TOKENIZER-TYPE=SMILES}}\token{MOLECULAR\_ENTITY}\token{MOLECULAR\_ENTITY\_SMALL\_MOLECULE}\token{SMILES\_SEQUENCE}\sequenceWrp{\texttt{CN(C)CCOc}\dots\texttt{[nH]2)cc1}} \tokenmeta{@TOKENIZER-TYPE=GENES}\token{MOLECULAR\_ENTITY}\token{MOLECULAR\_ENTITY\_CELL\_GENE\_EXPRESSION\_RANKED}\sequenceWrp{\texttt{[B2M][RPL10]\dots[ZBTB16][ZNF429]}}\token{EOS} & \metatoken{\token{@TOKENIZER-TYPE=SCALARS\_LITERALS}}\highlight{3.966226} \metatoken{\token{@TOKENIZER-TYPE=SMILES}} \dots \sequenceWrp{\texttt{[ZBTB16][ZNF429]}}\token{EOS} \\

Antibody design & \metatoken{\token{@TOKENIZER-TYPE=AA}}\complex{\token{COMPLEX\_ENTITY}}\molecule{\token{ATTRIBUTE\_ORGANISM}}\molecule{\token{ATTRIBUTE\_ORGANISM\_HUMAN}}\molecule{\token{MOLECULAR\_ENTITY}}\molecule{\token{MOLECULAR\_ENTITY\_TYPE\_ANTIBODY\_LIGHT\_CHAIN}}\sequenceWrp{\texttt{AB\dots CD}\highlight{\token{SENTINEL\_ID\_0}}\texttt{GF\dots KL}\highlight{\token{SENTINEL\_ID\_1}}\texttt{\dots TC}}\molecule{\token{MOLECULAR\_ENTITY}}\molecule{\token{MOLECULAR\_ENTITY\_TYPE\_ANTIBODY\_HEAVY\_CHAIN}}\sequenceWrp{\texttt{AA\dots DD}\highlight{\token{SENTINEL\_ID\_2}}\texttt{FK\dots KF}\highlight{\token{SENTINEL\_ID\_3}}\texttt{\dots JJ}}\molecule{\token{MOLECULAR\_ENTITY}}\molecule{\token{MOLECULAR\_ENTITY\_EPITOPE}}\sequenceWrp{\texttt{AB\dots GK}}\token{EOS} & \metatoken{\token{@TOKENIZER-TYPE=AA}}\highlight{\token{SENTINEL\_ID\_0}ABC}\highlight{\token{SENTINEL\_ID\_1}DDDDD}\highlight{\token{SENTINEL\_ID\_2}EEFF}\dots\highlight{\token{SENTINEL\_ID\_3}GKGK}\token{EOS} \\

\taskher & \metatoken{\token{@TOKENIZER-TYPE=AA}} \globalent{\token{BINDING\_AFFINITY\_CLASS}\highlight{\token{SENTINEL\_ID\_0}}} \metatoken{\token{@TOKENIZER-TYPE=AA@MAX-LEN=700}} \molecule{\token{MOLECULAR\_ENTITY}\token{MOLECULAR\_ENTITY\_ANTIBODY\_HEAVY\_CHAIN}} \sequenceWrp{\texttt{EVQ...KSC}}\metatoken{\token{@TOKENIZER-TYPE=AA@MAX-LEN=700}} \molecule{\token{MOLECULAR\_ENTITY}\token{MOLECULAR\_ENTITY\_ANTIGEN}} \sequenceWrp{\texttt{MEL...YEG}}\token{EOS} & \token{@TOKENIZER-TYPE=AA}\highlight{\token{SENTINEL\_ID\_0}\token{1}}\token{EOS} \\

\tasktitan & \metatoken{\token{@TOKENIZER-TYPE=AA}} \globalent{\token{BINDING\_AFFINITY\_CLASS}\highlight{\token{SENTINEL\_ID\_0}}} \metatoken{\token{@TOKENIZER-TYPE=AA@MAX-LEN=700}} \molecule{\token{MOLECULAR\_ENTITY}\token{MOLECULAR\_ENTITY\_TCR\_BETA\_VDJ}} \sequenceWrp{\texttt{TIQ...TVV}} \metatoken{\token{@TOKENIZER-TYPE=AA@MAX-LEN=170}} \molecule{\token{MOLECULAR\_ENTITY}\token{MOLECULAR\_ENTITY\_EPITOPE}} \sequenceWrp{\texttt{LEPLVDLPI}} \token{EOS} & \token{@TOKENIZER-TYPE=AA}\highlight{\token{SENTINEL\_ID\_0}\token{1}}\token{EOS} \\

\taskddg & \metatoken{\token{@TOKENIZER-TYPE=AA}}\globalent{\token{GENERAL\_AFFINITY\_CLASS}}\metatoken{\token{@TOKENIZER-TYPE=SCALARS\_LITERALS}}\highlight{\textless MASK\textgreater} \metatoken{\token{@TOKENIZER-TYPE=AA}}\complex{\token{COMPLEX\_ENTITY}}\molecule{\token{MOLECULAR\_ENTITY}\token{MOLECULAR\_ENTITY\_GENERAL\_PROTEIN}}\sequenceWrp{\texttt{IS\dots VY}}\metatoken{\token{@TOKENIZER-TYPE=AA}}\complex{\token{COMPLEX\_ENTITY}}\molecule{\token{MOLECULAR\_ENTITY}\token{MOLECULAR\_ENTITY\_GENERAL\_PROTEIN}}\sequenceWrp{\texttt{DC\dots KCNF \dots KC}}\metatoken{\token{@TOKENIZER-TYPE=AA}}\token{MUTATED}\molecule{\token{MOLECULAR\_ENTITY}\token{MOLECULAR\_ENTITY\_GENERAL\_PROTEIN}}\sequenceWrp{\texttt{DC\dots KCQF \dots KC}}\token{EOS} & \metatoken{\token{@TOKENIZER-TYPE=AA}}\globalent{\token{GENERAL\_AFFINITY\_CLASS}}\metatoken{\token{@TOKENIZER-TYPE=SCALARS\_LITERALS}}\highlight{0.244}\metatoken{\token{@TOKENIZER-TYPE=AA}}\complex{\token{COMPLEX\_ENTITY}} \dots \sequenceWrp{\texttt{KC}}\token{EOS} \\

DTI & \metatoken{\token{@TOKENIZER-TYPE=AA}}\highlight{\textless MASK\textgreater} \metatoken{\token{@TOKENIZER-TYPE=AA}}\token{MOLECULAR\_ENTITY} \token{MOLECULAR\_ENTITY\_GENERAL\_PROTEIN} \sequenceWrp{CC=...\metatoken{\token{@TOKENIZER-TYPE=SMILES}}\molecule{\token{MOLECULAR\_ENTITY} \token{MOLECULAR\_ENTITY\_SMALL\_MOLECULE}}}\sequenceWrp{AD...}\token{EOS} & \metatoken{\token{@TOKENIZER-TYPE=SCALARS\_LITERALS}} \highlight{\{standardized pKd\}} \\

\caption{ Examples (per fine-tune task) of encoder input ("prompt") and Decoder label (ground truth labels). "..." is used when part of the sequence text is not shown, for brevity and readability. For example "QG...SF" instead of "QGCQVVQGNLELTYLPTNASLSF" }
\end{longtable}
\endgroup

\begin{table}[H]
\caption{Cell type additional results}
\label{table_cell_type_results}

\begin{tabular}{lllll}
model & Accuracy & F1 & Precision & Recall \\
\midrule
scBERT & 0.759     & 0.691     & N/A & N/A \\
CIForm & 0.820     & 0.710     & N/A & N/A \\
MAMMAL & \textbf{0.856±0.004} & \textbf{0.763±0.012} & 0.774±0.016 & 0.761±0.011 \\
\end{tabular}
\end{table}

\begin{table}[H]
\caption{Statistics of GDSC Datasets}
\label{table_gdsc_data_stats}

\begin{tabular}{lM{2.5cm}M{2.5cm}M{2.5cm}}
Dataset & \# Cell lines & \# Drugs & \# Cell-Drug pairs \\
\midrule
 Cancer-Drug Resp.1 & 958  & 208 & ~177K \\
 Cancer-Drug Resp.2 & 805  & 137 & ~92K  \\
 Cancer-Drug Resp.3 & 561  & 223 & ~107K \\
\end{tabular}
\end{table}

\begin{table}[H]
\caption{Ab Infilling additional results}
\label{table_antibody_design_results}
\begin{tabular}{L{1.5cm}R{1.2cm}R{1.2cm}R{1.2cm}R{1.2cm}R{1.2cm}R{1.2cm}}
model & CDRH1-AAR & CDRH2-AAR & CDRH3-AAR & CDRL1-AAR & CDRL2-AAR & CDRL3-AAR \\
\midrule
dyMEAN \cite{kong2023end} & 0.757     & 0.685     & 0.375     & 0.755     & 0.831     & 0.521     \\
MAMMAL & \textbf{0.832 ±0.003} & \textbf{0.742 ±0.012} & \textbf{0.446 ±0.002} & \textbf{0.780 ±0.017} & \textbf{0.844 ±0.012} & \textbf{0.724 ±0.010} \\
\end{tabular}
\end{table}

\newpage
\section{Additional results - AlphaFold3}\label{appendix_alphafold3}

\begin{table}[H]
\caption{Performance comparison between MAMMAL (pre-trained and fine-tuned) and various AlphaFold3 (AF3) configurations for antibody-antigen binding prediction. The metrics are Area Under the Receiver Operating Characteristic curve (AUROC) and Area Under the Precision-Recall Curve (AUPRC)
}
\label{table_AF3_HER2}
\begin{tabular}{L{6cm}L{3cm}L{2.8cm}}

Method & AUROC & AUPRC \\
\midrule
MAMMAL Pre Trained & 0.491 & 0.473 \\
MAMMAL Fine Tuned & \textbf{0.931} & \textbf{0.928} \\
AF3 Heavy chain only ipTM & 0.448 & 0.489 \\
AF3 Heavy chain only pTM & 0.496 & 0.514 \\
AF3 Heavy+Light chains ipTM & 0.504 & 0.519 \\
AF3 Heavy+Light chains pTM & 0.497 & 0.519 \\
Control: AF3 H CD206 ipTM & 0.455 & 0.5 \\
Control: AF3 H CD206 pTM & 0.66 & 0.583 \\

\end{tabular}
\end{table}

\begin{table}[H]
\caption{Performance comparison of nanobody-antigen binding prediction on three targets measured by AUPRC and AUROC, between MAMMAL (pre-trained and fine-tuned) and AF3.}
\label{table_AF3_nanobody}
\begin{tabular}{L{5cm}|lll|lll|}
Method & \multicolumn{3}{c|}{AUPRC $\uparrow$} & \multicolumn{3}{c|}{AUROC $\uparrow$} \\
 & CD206 & VWF & TBG & CD206 & VWF & TBG \\
\midrule
MAMMAL Pre Trained & 0.359 & 0.464 & 0.376 & 0.478 & 0.505 & 0.425 \\
MAMMAL Fine Tuned & 0.995 & 0.802 & 0.526 & 0.997 & 0.832 & 0.700 \\
AF3 ipTM & 0.434 & 0.358 & 0.953 & 0.565 & 0.402 & 0.976 \\
AF3 pTM  & 0.526 & 0.318 & 0.942 & 0.651 & 0.552 & 0.964 \\
\end{tabular}
\end{table}

\begin{figure}[h]
\centering
\includegraphics[width = 350pt]{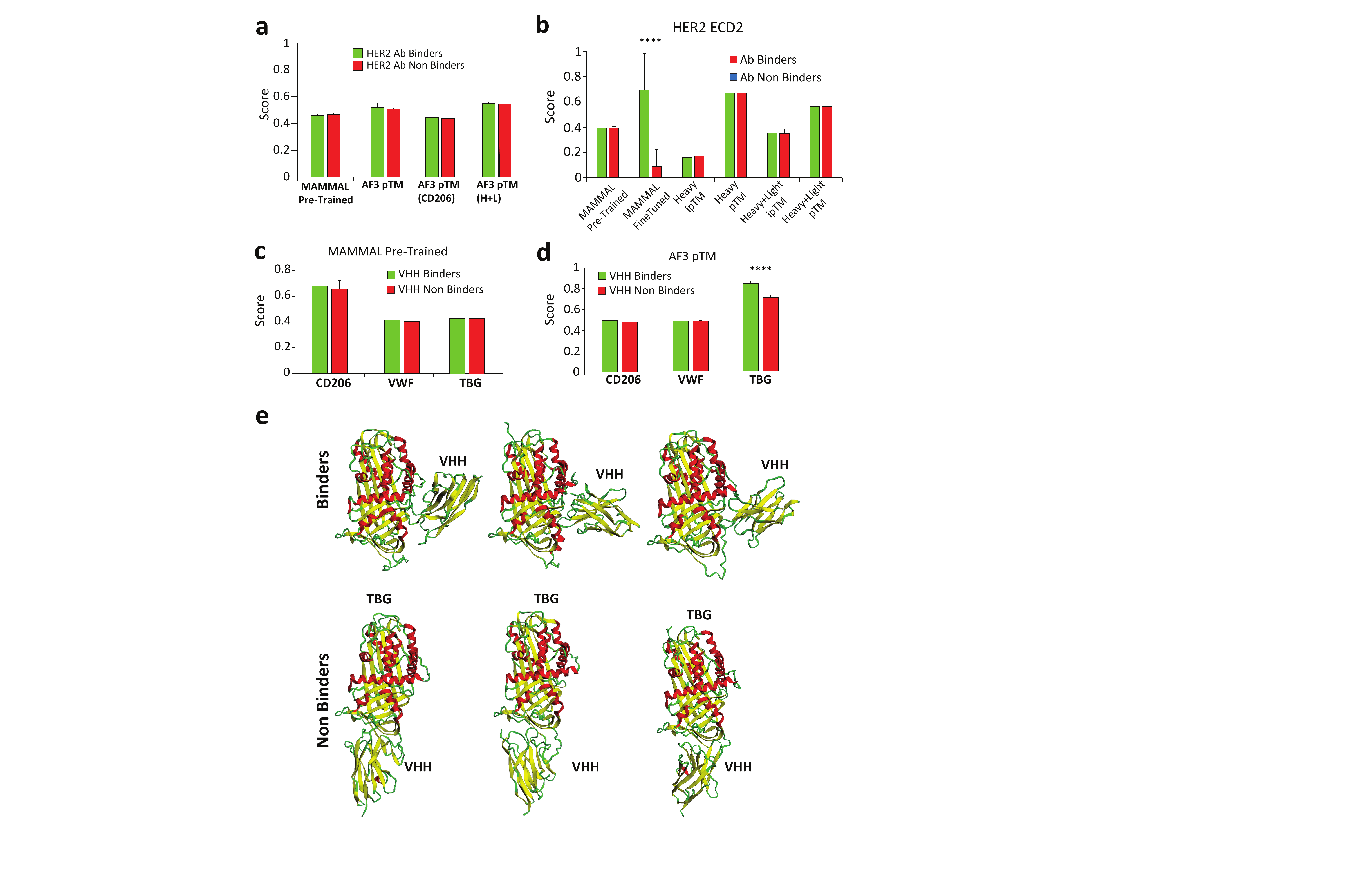}
\caption{\textbf{(a)} Evaluation of binding to an alternative HER2 representing ECD sequence and antibodies (heavy alone or with light chain). CD206 shown as control. pTM shown for completeness in addition to ipTM. \textbf{(b)} Testing with an alternative HER2 ECD (Uniprot P04626) \textbf{(c)} Testing with a pretrained version of MAMMAL which was not fine-tuned on an antigen-antibody binding task, on antigen-antibody binding prediction \textbf{(d)} Testing AF3 pTM scores on antigen-nanobody binding prediction \textbf{(e)} TBG and three binding and three non-binding VHHs (nanobodies) as predicted by AF3. }\label{fig_sup_af3}
\end{figure}

\FloatBarrier

\end{appendices}

\end{document}